\documentclass[prd,
               twocolumn,
               eqsecnum,
               fixfloats,aps,amsmath,amssymb,
               showpacs,
               letterpaper,
               superscriptaddress,
               altaffilletter,
               nofootinbib, nobibnotes
               ]{revtex4-1}

\usepackage{enumerate}
\usepackage{enumitem}
\usepackage{mathrsfs}
\usepackage{wasysym}
\usepackage{multirow}
\usepackage[pdftex]{graphicx}
\usepackage[plainpages=false]{hyperref}
\usepackage{epstopdf}
\usepackage{longtable}
\usepackage[export]{adjustbox}[2011/08/13]
 
\newcommand{ \regtext }[1]{\mbox{\tiny#1}}

\begin{document}


\title{Measuring violations of General Relativity from single gravitational wave detection by non-spinning binary systems: higher-order asymptotic analysis}

\author{Rhondale Tso}\thanks{rtso@ligo.caltech.edu}
\affiliation{LIGO Laboratory, California Institute of Technology, Pasadena, CA 91125, USA}
\affiliation{Columbia Astrophysics Laboratory, Columbia University, New York, NY 10027, USA}
\affiliation{Physics Department, Embry-Riddle Aeronautical University, Prescott, AZ 86301, USA}

\author{Michele Zanolin}\thanks{zanolin@ligo.mit.edu}
\affiliation{Physics Department, Embry-Riddle Aeronautical University, Prescott, AZ 86301, USA}

\date{\today}

\begin{abstract}
A frequentist asymptotic expansion method for error estimation is employed for a network of gravitational wave detectors to assess the amount of information that can be extracted from gravitational wave observations.  Mathematically we derive lower bounds in the errors that any parameter estimator will have in the absence of prior knowledge to distinguish between the post-Einsteinian (ppE) description of coalescing binary systems and that of general relativity.  When such errors are smaller than the parameter value, there is possibility to detect these violations from GR.  A parameter space with  inclusion of dominant dephasing ppE parameters $(\beta, b)$ is used for a study of first- and second-order (co)variance expansions, focusing on the inspiral stage of a nonspinning binary system of zero eccentricity detectible through Adv.\,LIGO and Adv.\,Virgo.  Our procedure is an improvement of the Cram\'{e}r-Rao Lower Bound.   When Bayesian errors are lower than our bound it means that they depend critically on the priors.  The analysis indicates the possibility of constraining deviations from GR in inspiral SNR ($\rho \sim 15-17$) regimes that are achievable in upcoming scientific runs (GW150914 had an inspiral SNR $\sim 12$).  The errors on $\beta$ also increase errors of other parameters such as the chirp mass $\mathcal{M}$ and symmetric mass ratio $\eta$.  Application is done to existing alternative theories of gravity, which include modified dispersion relation of the waveform, non-spinning models of quadratic modified gravity, and dipole gravitational radiation (i.e., Brans-Dicke type) modifications.
\end{abstract}


\maketitle


\section{Introduction}

The advanced generation of the LIGO-Virgo network of interferometers~\cite{ADV_LIGO, LIGO, VIRGO} started collecting data in September 2015 and provided the first detection of gravitational waves (GWs)~\cite{det_paper}, allowing to start testing General Relativity (GR) beyond current constraints~\cite{Will_grav_LRR} into strongly relativistic regimes~\cite{test_gr, Gair_et_al_LRR, Yunes_Siem_2013}.  In this paper we quantify the capability of laser interferometers to detect violations of GR, with a single detection of a compact binary coalescence  signal, by assessing if the minimal error on the  parameterized post-Einsteinian (ppE) parameters are larger than the separation of modified gravity values with respect to standard GR values.  Error bounds are computed with the most accurate frequentist approach to date by computing the errors as inverse power series in the signal-to-noise ratio (SNR), where the first order is the inverse of the Fisher information matrix~\cite{Vit_Zan_Mak_2010_single, Vit_Zan2010_BeyondFisher, Vit_Zan2011_Net}.  In this paper we model GR violations with the ppE framework~\cite{Yun_Pret2009_FundBias,Yunes_Siem_2013,Arun_et_al_2006, Arun_et_al_2006_2,  Arun_et_al_2010}, which produces parametrized extensions of GR GW signals for the inspiral phase only of a binary compact coalescence in the absence of spin (similar extensions are currently not available for the merger and ringdown phase as well as in the presence of spin).  

The square root of the inverse Fisher matrix diagonal elements, also known as the Cram\'{e}r-Rao Lower Bound (CRLB), is a lower limit in the error of any unbiased estimator in the absence of prior knowledge. In this regard the CRLB is a statement about the amount of information available in the data regardless of the specific parameter estimation scheme.  There is however no guarantee that any estimator is capable to actually attain the CRLB for part or the whole range of values the physical parameters can assume. Also, the CRLB only takes into account the curvature of the probability distribution of the data around the true value of the parameters and therefore does not include the role of secondary maxima in the calculation of the variance or mean square error of the estimators. The improved bound adopted here (based on second order asymptotics) is larger than the inverse Fisher matrices, known to underestimate errors in low-SNR detections. Second-order bounds have been previously used for compact binary coalescence waveforms in quantifying the accuracy in intrinsic parameters as well as the direction of arrival for a network of laser interferometers \cite{Vit_Zan_Mak_2010_single, Vit_Zan2010_BeyondFisher, Vit_Zan2011_Net}. 

The benefits of using the second order of the expansions is in the fact that they depend up to the fourth derivative of the likelihood function and, therefore, are sensitive to asymmetries and side lobes of the estimator probability distribution (similar to the change in the accuracy of a Taylor expansion when extended to higher orders).  Also, in the past~\cite{Vit_Zan_Mak_2010_single, Vit_Zan2010_BeyondFisher, Vit_Zan2011_Net}, the comparison of the second order with the first order provided an analytical understanding of the reasons the CRLB could not be met (for example, in Ref.~\cite{Vit_Zan_Mak_2010_single}, a novel relationship between the Kurtosis of the probability distribution of the estimator and the SNR was derived to understand when the CRLB could be met).

Bayesian methods were recently applied to test modified GR  signals through consistency tests~\cite{Poz_et_al_2012, Poz_et_al_2012_2}, and the ppE framework~\cite{Corn_Samp_Yun_Pret2011_GWtests}.  Refs.~\cite{Poz_et_al_2012, Poz_et_al_2012_2} developed a framework to detect GR violations without modeling the violation, this works in the limit of large number of detections.  This framework was used in GR tests from the GW150914 transient~\cite{test_gr}.  Bayesian  selection methods were also used in Ref.~\cite{Corn_Samp_Yun_Pret2011_GWtests} and  Ref.~\cite{Pozzo_Veitch_Vecchio2011} to constrain the range of ppE parameter values, provided that priors are adopted.  

When Bayesian uncertainties are smaller than the frequentist bounds, it means that the parameter estimation errors depend critically on the priors.  This issue can be an artifact if the prior is not based on previous detections or no robustness studies were performed with respect to the choice of the priors  (see for example the discussion in Ref.~\cite{Sivia} of the effects of priors).  In this paper, we show that this instance happens for an equal-mass binary black hole system in the massive graviton case.  This example illustrates how the present work provides a unique understanding of the parameter estimation errors.  Although GW150914 had a SNR$\sim 24$, its inspiral stage falls within the prescribed study of SNR$ < 20$.~\footnote{GW150914 has inspiral SNR$\sim 12$.}  

In addition, this work extends the Fisher information based results of Ref.~\cite{Arun_et_al_2006, Arun_et_al_2006_2,  Arun_et_al_2010}, which perform error estimations by modifying PN coefficients.  We also extend Fisher-based assessments of specific alternative theories~\cite{Will_1994_BD, Will_1998_MG, Arun_Will_2009, Kep_Ajith2010_MG_Bound, Mirs-Yunes-Will_2012_LV}.  Specifically, this paper considers phase modification in  the {\it restricted} ppE framework~\cite{Yunes_Siem_2013}, considering the ppE framework as a general enhancement to existing {\texttt{TaylorF2}}~\cite{taylorf2, Arun_et_al_2005} GR templates in a three detector LIGO-Virgo network~\cite{ADV_LIGO, LIGO, VIRGO}.   Calculations in this limit were chosen since deformations to the GW's phase are expected to be more resolvable~\cite{Corn_Samp_Yun_Pret2011_GWtests, Apostolatos_FF} and complements recent Bayesian methods testing deviations from GR~\cite{Poz_et_al_2012, Poz_et_al_2012_2}.   Second-order frequentist constraints produced in this paper are at the same order of magnitude as  the Bayesian model selection's errors in Ref.~\cite{Corn_Samp_Yun_Pret2011_GWtests}, where our errors are quantified at the one sigma level.  As error estimates of  ppE parameters grow, second-order errors of  parameters such as the chirp mass, symmetric mass ratio, and time of coalescence also inflate.   The results presented here, and the rescaled bonds which can be simply derived by changing the SNR, will be important benchmarks for any parameter estimation scheme which will be used in existing and future interferometer data, including Bayesian parameter estimation algorithms.

Section~\ref{stat_model_sig} of this paper introduces the signal model used.
Section~\ref{param_space_expan} discusses what is the resolvable parameter space and the expansion model, in particular subsection~\ref{ppE_model} discusses alternative theories of gravity covered in this paper and the asymptotic expansion of the maximum likelihood estimator model is discussed in subsection~\ref{expan}.
Finally, section~\ref{results} assesses the results, as applied to a two-dimensional ppE parameter space (\ref{bidimen}) and a seven-dimensional parameter space of equal mass (\ref{7dimen}) and unequal mass (\ref{7dimen2}) systems with  physical parameters included.
Results are applied to existing alternative theories of gravity in~\ref{mod_grav}, including massive graviton, Brans-Dicke, and quadratic modified gravity (encompassing Einstein-Dilation-Gauss-Bonnet gravity).
A summary and discussion is given in section~\ref{conclusion}.  


\subsection{Signal Model}\label{stat_model_sig}

The waveforms are assumed to be produced by a nonspinning binary system with all orbital eccentricity information lost when entering the frequency bandwidth of Adv.\,LIGO and Adv.\,Virgo.  Fourier transform of the signal, through stationary phase~\cite{filters_gw_binaries, bender_orszag}, becomes,
\begin{equation}\label{freq_sig_GR}
s_{\regtext{GR}}^I (f) = A_{\regtext{GR}}^I(f) e^{i \left(  \psi_{\regtext{GR}}(f) - 2 \pi f \tau_I  - \Phi_0^I \right) } \, , \,\,\,\, f < f_{\regtext{merg}}
\end{equation}
for the inspiral stage of the compact binaries. For the phase  $\psi_{\regtext{GR}}(f)$ and amplitude $A_{\regtext{GR}}^I(f)$ the standard {\texttt{TaylorF2}} model~\cite{taylorf2, Arun_et_al_2005} is used.

The signal of a collection of alternative theories of gravity is modelled as~(\ref{freq_sig_GR}) modulated in the phase and amplitude as:
\begin{eqnarray}\label{modified_gr}
A^I_{\regtext{GR}}(f) &\rightarrow& A^I_{\regtext{GR}}(f) \left( 1 + \delta A(f) \right) ,  \\
\psi_{\regtext{GR}} (f) &\rightarrow& \psi_{\regtext{GR}} (f) + \delta \psi (f), \nonumber
\end{eqnarray}
where  $\delta A(f)$ and $\delta \psi (f)$ are a general series of  scaling parameters $\alpha_i, \beta_i \in \Re$ and in some instances arguments call for integer exponentials of  $\nu \eta^{1/5}$~\cite{Vallisneri_Yunes_stealth_bias, Chat_Yun_Corn2012_ExtendedppE}, where $\nu = (\pi M f)^{1/3}$ for total mass $M$ and $\eta = m_1 m_2 / M^2$.  Here  the analysis is done at leading order in the ppE parameters,
\begin{eqnarray}\label{lead_order}
\delta A_{\regtext{ppE}}(f) &=& \alpha(\nu \eta^{1/5})^{a},\\
 \delta \psi_{\regtext{ppE} } (f) &=& \beta(\nu \eta^{1/5})^{b}, \nonumber
\end{eqnarray}

At each interferometer the signal is assumed to be recorded with additive noise as in Ref.~\cite{Vit_Zan2011_Net}.  Frequency dependent noise for Adv.\,LIGO are interpolated from the official power spectral density~\cite{LIGO_noise} of high-power, zero-detuning.  Adv.\,Virgo  is assumed to have the sensitivity given in Ref.~\cite{VIRGO_noise}.  For error analysis, and upcoming integrations, the lower cutoff frequency is set to $f_{\regtext{low}}$ and the upper cutoff is set to the upper limit for reliability in the inspiral of the waveform template, i.e., the innermost stable circular orbit (ISCO) frequency,
$$ f_{\regtext{low}} = 20 \mbox{ Hz}\,\, , \,\,\,\, f_{\regtext{up}} = f_{\regtext{ISCO}} \approx (6^{3/2} \pi M )^{-1}.$$
The convention used in~(\ref{freq_sig_GR}) is presented in Appendix~\ref{appendA}.

\section{Parameter space and Expansion}\label{param_space_expan}

For  non-spinning systems thirteen parameters are necessary in the description of the inspiral of two coalescing binaries: two mass terms, four angles (two source location and two waveform angles), two coalescence parameters, distance to the source, and four ppE parameters in the leading order approximation.  Singular Fisher matrices might appear~\cite{Vit_Zan_Mak_2010_single, use_abuse_fisher}, indicating that the resolvable parameter space is smaller (where the Fisher matrix approach can still be used).

The distance $D_L$ is excluded from the error estimates because the amplitude has a dependency on both mass and distance parameters, and the independent treatment of both is unresolvable as already indicated in Ref.~\cite{Vit_Zan2011_Net}.  The coalescence phase is also not included because estimations of $\phi_c$ is relevant only when a full waveform (inspiral, merger, and ringdown) is implemented.  The polarization $\psi$ is excluded because results tend to be independent of it~\cite{Vit_Zan2011_Net}.  

\begin{figure}
\begin{centering}
\includegraphics[scale=0.35]{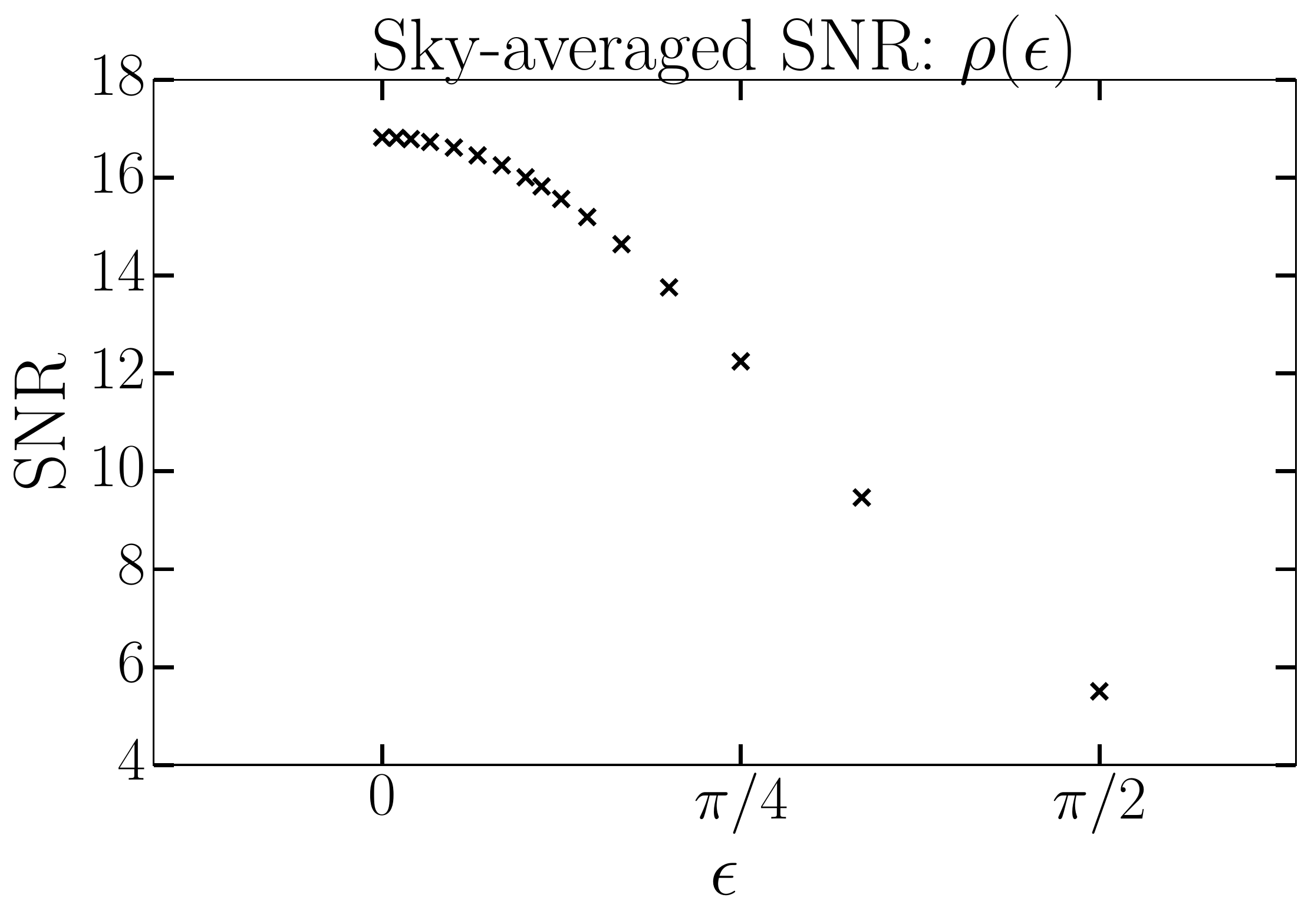}
\includegraphics[scale=0.35]{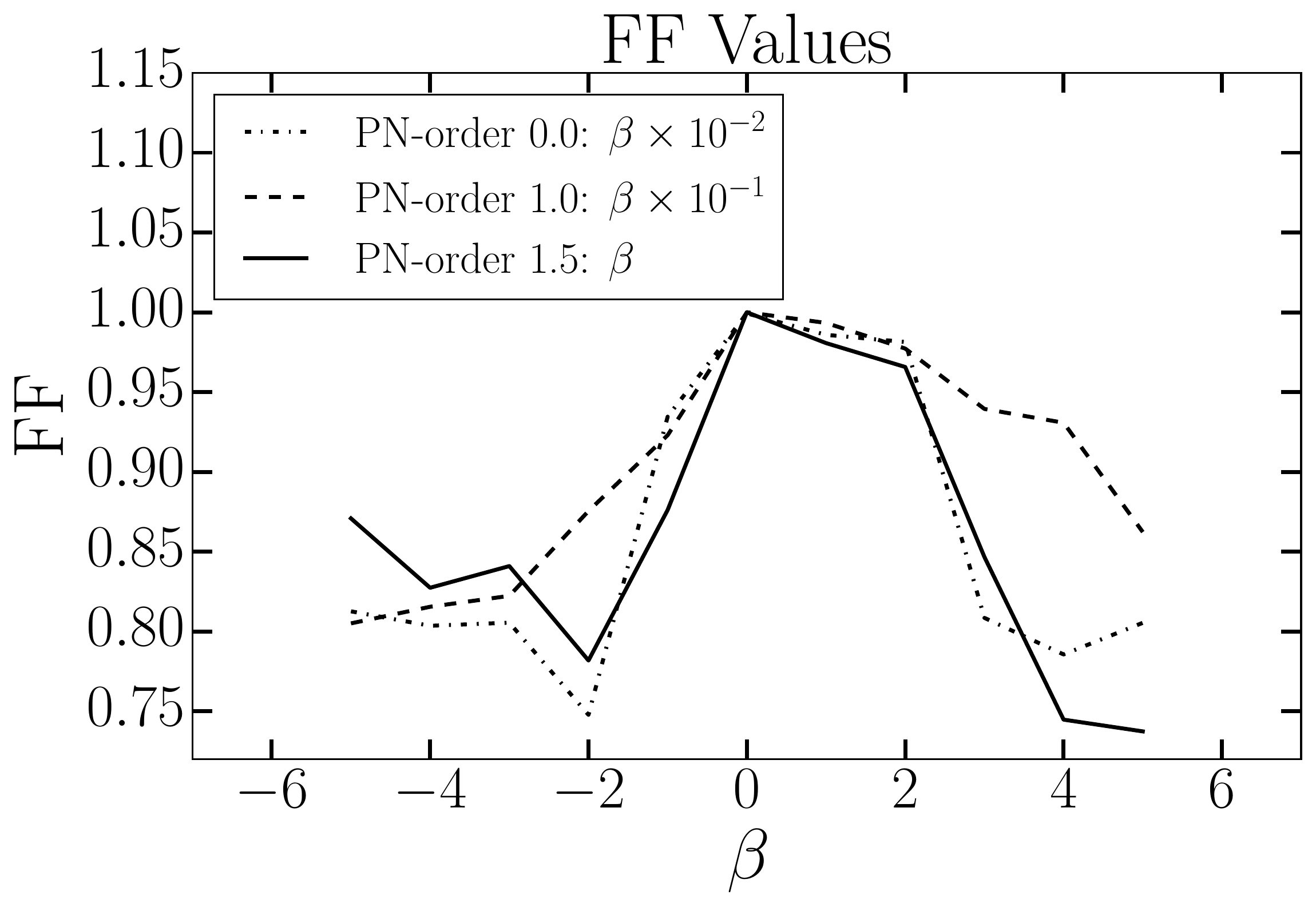}
\end{centering}
\caption{
Top: Sky-averaged SNR plotted with $\epsilon$ varied for system parameters: $m_1 = m_2 = 10 M_{\odot}, t_a = \phi_a = 0, \beta = -0.2, D_L = 1100$ Mpc, and $b=-3$ in the three detector network.
Bottom: Fitting factors~(\ref{ff}) for a range of $\beta$ with $b$ fixed to produce PN-order 0.0, 1.0, and 1.5 modifications for a system of: $m_1 = m_2 = 10 M_{\odot}$ and $t_a = \phi_a = 0$.  Adv.\,LIGO noise is assumed.  Since the range of $\beta$-values scale differently at each PN-order, each $\beta$-interval is scaled (as labeled in the legend).  For example, in the PN-order 0.0 modification the $\beta$ values in the domain are each scaled by $10^{-2}$.
}
\label{fitting_factor}
\end{figure}

Derivatives of the fitting factor $(FF)$~\cite{Apostolatos_FF},
\begin{equation}\label{ff}
FF =  \max_{ \vec{\zeta} }  \left( \frac{ \langle s_1 (\vec{\lambda}) | s_2 (\vec{\zeta}) \rangle }{ \sqrt{ \langle s_1 (\vec{\lambda}) | s_1 (\vec{\lambda}) \rangle} \sqrt{\langle s_2 (\vec{\zeta}) | s_2 (\vec{\zeta})  \rangle}  } \right)
\end{equation}
with respect to the binary's inclination $\epsilon$ evaluated at, or in a neighborhood of, $\epsilon = 0$ are roughly zero leading to impossibility to estimate $\epsilon$ and  singular Fisher matrices.  Here the $\langle \cdot | \cdot \rangle$ represent noise weighted inner products~\cite{Vit_Zan2011_Net, maggiore} and  $s_{1,2}$  are GW signals controlled by general parameter space vectors $\vec{\lambda}$ and $\vec{\zeta}$.  Keeping other parameters fixed and varying only $\epsilon$  produces change in the SNR equivalent to the rescaling of the distance, which affects GW plus-cross polarizations similarly.  Top panel in figure~\ref{fitting_factor} shows the sky-averaged SNR plotted as a function of inclination $\epsilon$ (only the GR polarizations are considered). Also, sky patterns of the errors  remain consistent when varying $\epsilon$.  Therefore, since $\epsilon$ is degenerate with $D_L$ it is also excluded from our resolvable parameter space, which becomes  $\theta_{\regtext{phys}}^i = \{ \eta, \log \mathcal{M}, t_c, \mbox{lat}, \mbox{long}  \}$.

Throughout this paper amplitude modulations are to be held fixed to that of GR: $\alpha = 0$, because the same effect could be produced by changing physical parameters like distance or mass.   Such an approach supposes that GR-violating amplitudes in the waveform are suppressed or modifications manifest only in waveform propagation.\footnote{Modifications to just propagation could surface through alterations in the dispersion of the GW, with alterations stemming from waveform generation excluded~\cite{Will_1998_MG, Mirs-Yunes-Will_2012_LV}.  Past studies also indicate modulations are most sensitive to phase modulations~\cite{Corn_Samp_Yun_Pret2011_GWtests, Apostolatos_FF}.}  Also, recent work suggests that GR modifications produced during the generation of a  waveform can be disentangled from that produced during propagation~\cite{Chat_Yun_Corn2012_ExtendedppE}, thus, in the event that phase deformation dominates GR-violating effects, amplitude modifications can be disregarded.  Calculations in this restricted framework are performed with modifications at various PN-orders in the phase, where in the strong-field regime discrete values of $b$ controls what PN-order correction is constituted for free parameter $\beta$ (GR result: $\beta=0$).

A qualitative way to study the influence of ppE parameters $(\beta, b)$ on a GR signal can be obtained through the correlation of the signals by means of the fitting factor~(\ref{ff}).  Each integration is done from 20 Hz to $f_{\regtext{ISCO}}$ with the noise curve of Adv.\,LIGO~\cite{LIGO_noise} ``high-power-zero-detuning."  Our exact waveform $s_1$ is represented by a {\texttt{TaylorF2}} waveform, whereas, a modified-{\texttt{TaylorF2}},  formed through~(\ref{modified_gr}) and~(\ref{lead_order}), acts as  $s_2$.  So $\vec{\lambda}$ is  the GR-limit parameter space vector and $\vec{\zeta}$ is that of the ppE parameter space.  The inner products are  maximized over evenly spaced parameters $\vec{\zeta}$ to provide a $FF$-value, where $FF=1$ represents an exact match between signals. Both {\texttt{TaylorF2}} models are kept to PN-order 3.5 in the phase.  In the denominator of~(\ref{ff}), amplitude parameters normalize to leave $f^{-7/3} / S_h$ in each integrand.  The numerator retains integrand $( f^{-7/3} / S_h ) e^{i  \Delta \psi(f;  \vec{\lambda}, \vec{\zeta})}$, where,
$$ \Delta \psi(f;  \vec{\lambda}, \vec{\zeta}) = \psi(f; \vec{\lambda}) - \psi(f;\vec{\zeta}) -  \delta \psi_{\regtext{ppE} } (f) $$
 and, in fixing $b$ and varying $\beta$, the parameters needing to be maximized over are $\vec{\zeta} = \{ t_c , \phi_c, \eta, M_{\regtext{tot} }  \}$.  Parameters are evenly spaced, in a $30\times30\times 30\times 30$ grid, within intervals: $0.05 \leq \eta \leq 0.25$, $0.5 M_{\regtext{tot}} \leq M_{\regtext{tot}} \leq 1.5 M_{\regtext{tot}}, -\pi \leq \phi_c \leq \pi$, and $-1.3\times 10^{-2} \leq t_c \leq 1.3\times 10^{-2}$.
 
Figure~\ref{fitting_factor} displays the results for an equal-mass system of $m_1 = m_2 = 10 M_{\odot}$ and $t_a = \phi_a = 0$ for PN-order 0.0, 1.0, and 1.5 modifications in the waveform.  Parameters $\vec{\zeta}$ are maximized over for a variety of $\beta$-values.  Note that at lower PN-orders the interval of $\beta$ is scaled differently than the $-5 \leq \beta \leq 5$ depicted, an interval valid for PN-order 1.5 modifications. The general trend is that the fitting factor is less affected by $\beta$ for larger PN-order with a skew in the $FF$-distribution towards the positive domain of $\beta$-values.

\subsection{Restricted ppE template and existing dephasing alternatives}\label{ppE_model}

As stated, variations of $\beta$ are restricted to fixed PN-order corrections in the phase.  For the two-dimensional study $b$ is fixed to induce modifications at (separately) PN-orders 0.0, 0.5, 1.0, 1.5, 2.0, and 3.0 which acts as a demonstration to the error estimation procedure.  Higher-dimensional studies specifically target a PN-order 1.0 modification and a weak-field $b=-7$ modification to address dispersion modification and dipole gravitational radiation.  From this reason $\beta$ is varied with error estimations performed at each $\beta$-value.  In Ref.~\cite{binary_pulsar_ppE} an analysis of binary pulsar PSR J0737-3039~\cite{binary_pulsar} placed bounds on ppE parameters (for this binary $4\eta \approx 1$ as determined from radio pulsar measurements~\cite{binary_pulsar}).  At PN-order 2.5 ($b=0$) degeneracies occur with other fiducial parameters, thus is not considered in the analysis.  In some theories constraints for $b=-7$ cannot be implemented from pulsar measurements, due to $\beta$'s dependence on mass differences of the system and other  theoretical parameters which will be discussed shortly.  With the exception of $b=-7$, parameters that probe weak-field ($b<-5$) are not considered since they are better constrained via binary pulsar measurements~\cite{Corn_Samp_Yun_Pret2011_GWtests}.  

At $b=-7$, the even-parity sector of quadratic modified gravity (QMG), an example being Einstein-Dilation-Gauss-Bonnet (EDGB) gravity, can be explored. For even-parity QMG, the violating term for a BBH system depends on the mass differences of the BHs: $\beta \propto \zeta_3 \eta^{-18/5} (1-4 \eta)$, unresolvable for equal-mass systems~\cite{QMG}.  For BHNS systems, the violating coefficients depend on the ratio of the two bodies: $\beta \propto \zeta_3 \eta^{-8/5} (m_{\regtext{NS}} / m_{\regtext{BH}})^2$ due to the `scalar charge' vanishing in NSs~\cite{QMG, xray_binary}.  With this same $b=-7$ correction, examples of dipole gravitational radiation, like Brans-Dicke (BD), can also be assessed.  Here BD-like modifications further depend on the difference of parameters which measure the body's inertial mass variations with respect to the local background value of the effective gravitational constant.  These so-called `sensitivity parameters' $s_{\regtext{BH,NS}}$ are generally set to 0.5 for black holes, so their difference vanish for a BBH system.  Only a BHNS system would allow constraints of BD-like modifications since $0.2 \leq s_{\regtext{NS}} \leq 0.3$~\cite{Healy_2012, Mirshekari_Will_2013, Yunes_Pani_Cardoso_2012, Zaglauer_1992}. 

For corrections at $b \neq -7$, most existing modifying coefficients depend on parameters that either vanish in the non-spinning model~(\ref{freq_sig_GR}) or contribute beyond PN-order 3.5.  This is the case in specific models of QMG, e.g., the odd-parity sector and dynamical Chern-Simons (CS) gravity~\cite{QMG}.  As an example, in the circular inspiral of two comparable mass BHs the GR-deviating term of dynamical CS has dependencies on the BH spins $\hat{S}_{1,2}$ and their relations to their orbital angular momentum $\hat{L}$: $\delta C = \delta C (m_{1,2}, \hat{S}_{1,2}, \hat{L})$~\cite{Dyn_CS}.  When the binary system is non-spinning, modifications are beyond PN-order 3.5. 

Beyond modifications during waveform generation, two propagating effects are massive graviton (MG) and simplified versions of Lorentz-violating (LV) theories~\cite{Will_1998_MG, Mirs-Yunes-Will_2012_LV}.   Parameters to constrain are the graviton Compton wavelength $\lambda_g$ and $\lambda_{\regtext{LV}} = 2 \pi \mathbb{A}^{1/(\gamma - 2)}$.  Here $\mathbb{A}$ is a phenomenological parameter modifying the gravitational waveform's dispersion relation.   The $\gamma$-dependent distance measure $D_{\gamma}$ (see Ref.~\cite{Mirs-Yunes-Will_2012_LV} for exact formula) further depends on known astrophysical parameters (Hubble parameter, matter density parameter, etc.) which are assumed to be exact knowns in the analysis~\cite{PDG_2012}.  Parameter $\gamma$ governs the order of correction and $\gamma =0$ (PN-order 1.0) is what we're limited to since this is the only value contained in the ppE framework for the PN-order 3.5 {\texttt{TaylorF2}} model.  Such MG-LV interpretations  are generic models modifying the dispersion of a GW with more specific generation mechanism still yet to be explored.   Ref.~\cite{Poz_et_al_2012} notes some limitations in prescribing MG effects as modifications of the dispersion of the waveform.  In LV-type modification further work in existing, model-independent approaches, e.g., the Standard Model Extension~\cite{kostelec, kostelec_2004}, could be interesting (see for example Ref.~\cite{kostelec_2015}).
 
Constraints have been imposed on the wavelength of the graviton.  The detection of GW150914 and binary-pulsar constraint serve as dynamical bounds while solar-system constraints, serving as static bounds, provide the most reliable estimates~\cite{test_gr, Bert-Gair-Ses_2011_MG}.  So, parameters are represented by,
\begin{eqnarray}
\lambda_{\regtext{LV}}= 2 \pi \mathbb{A}^{-1/2},
\,\,\,\,\,\, 
\lambda_g \geq
\begin{cases}
 10^{13} [\mbox{km}], & \mbox{dynamic (GW)}, \\
1.6 \times 10^{10} [\mbox{km}], & \mbox{dynamic (pulsars)}, \\
2.8 \times 10^{12} [\mbox{km}], & \mbox{static}. 
\end{cases} \nonumber
\end{eqnarray}
For EDGB  gravity, the constraint parameter is $|\alpha_{\regtext{EDGB}}|$.  Here  $\zeta_3 = \xi_3 M^{-4} =   16 \pi  \alpha^2_{\regtext{EDGB}}  M^{-4}$, with $\beta_{\regtext{BBH}} \propto \zeta_3 \eta^{-18/5} (1-4 \eta)$ and $\beta_{\regtext{BHNS}} \propto \zeta_3 \eta^{-8/5}(m_{\regtext{NS}} / m_{\regtext{BH}})^2$.  In Brans-Dicke theory $\beta \propto (s_{\regtext{BH,NS}} - s_{\regtext{BH,NS}} )^2 \omega_{\regtext{BD}}^{-1}$.  From measurements  of the Cassini spacecraft~\cite{cassini, cassini2}  bounds on EDGB and Brans-Dicke parameters are,
\begin{eqnarray}
|\alpha_{\regtext{EDGB}}|^{1/2}  &\leq& 8.9 \times 10^{6} \,\, \mbox{km},	\nonumber \\
\omega_{\regtext{BD}}		  &>& 4 \times 10^{4} . \nonumber 
\end{eqnarray}
With other suggested constraints~\cite{xray_binary, EDGB_bound1} giving,
\begin{eqnarray}
|\alpha_{\regtext{EDGB}}|^{1/2}  &<& 1.9  \,\, \mbox{km}. \nonumber \\
|\alpha_{\regtext{EDGB}}|^{1/2}  &<& 9.8 \,\, \mbox{km}, \nonumber 
\end{eqnarray}
GW150914 results have allowed studies to infer the theoretical significance of the testing GR study~\cite{test_gr} in various specific models, see for example Refs.~\cite{test_gr_implications, test_gr_LV}.

\subsection{Asymptotic Expansions}\label{expan}
Similar to Ref.~\cite{Vit_Zan2011_Net}, we reasonably assume only Gaussian noise at time of the signal and that the noise is uncorrelated at different interferometers.  Here we use the analytic asymptotic expansion of the variance and bias developed in Refs.~\cite{Vit_Zan_Mak_2010_single, Vit_Zan2010_BeyondFisher, Vit_Zan2011_Net},
\begin{eqnarray}
\sigma^2_{\vartheta ^i} &=& \sigma^2_{\vartheta ^i}[1] + \sigma^2_{\vartheta ^i}[2] + \cdots, \label{variance} \\
b_{\vartheta ^i} &=& b_{\vartheta ^i}[1] + b_{\vartheta ^i}[2] + \cdots, \label{bias}
\end{eqnarray}
with $\sigma^2_{\vartheta ^j}$ being the diagonal elements of the covariance matrix, where
\begin{eqnarray}
 \sigma_{\vartheta ^j}[1] , b_{\vartheta ^j}[1] & \propto& \rho^{-1}, \nonumber \\
 \sigma_{\vartheta ^j}[2] , b_{\vartheta ^j}[2] & \propto& \rho^{-2}, \nonumber 
 \end{eqnarray}
for network SNR $\rho$.  This inverse proportionality continues at higher orders in similar fashion.  Here the network SNR is the sum over the square of the optimal SNR $\rho^I$ of the signal at the $I$-th detector,
\begin{equation}
\rho^2= \sum_{I} \left( \rho^I \right)^2 , \,\,\,\,  \rho^I  = \langle s^I | s^I\rangle^{1/2}
\end{equation}
Notice that $\rho$ increases for a fixed source by increasing the number of detectors.  The first-order term of the expansion of the variance, the diagonal components of the inverse Fisher matrix, dominates the bound on the error in the limit of large SNR, while higher order terms become more important for medium to low SNR.

What is usually regarded as the error in a lab measurement is the square root of the mean-squared error (MSE), where the MSE is the sum of the variance~(\ref{variance}) and square of the bias~(\ref{bias}): $MSE_{\vartheta ^i} = \sigma^2_{\vartheta ^i} + b^2_{\vartheta ^i}$.  Since this analysis computes errors at second-order of $1/\rho$, the expression above only requires first-order of the bias which is negligible as already discussed in Ref.~\cite{Vit_Zan2011_Net}. We estimate uncertainties of the two-dimensional ppE parameter space $\theta^i_{\regtext{ppE}}$ for different $\beta$ at a fixed exponential $b$.  In addition, the inclusion of $\theta^i_{\regtext{ppE}}$ to a signal's extrinsic and intrinsic parameter space $\theta^i_{\regtext{phys}}$ is also assessed.

Finally, error bounds are indicated with,
\begin{eqnarray}
\Delta \vartheta_i[1] &=& \sqrt{ \sigma_{\vartheta^i}^2[1] }, \phantom{scp}  \Delta \vartheta_i[2] = \sqrt{ \sigma_{\vartheta^i}^2[2] } \nonumber \\
\Delta \vartheta_i[1+2] &=&  \sqrt{\sigma_{\vartheta^i}^2[1] + \sigma_{\vartheta^i}^2[2]}.
\end{eqnarray}
For example first-order errors of the symmetric mass ratio $\eta$ are marked by $\Delta \eta[1]$, second-orders are marked by $\Delta \eta[2]$, and total error with the inclusion of second-order contributions as $\Delta \eta[1+2]$.

\begin{figure*}
\includegraphics[width=1.2\textwidth,center]{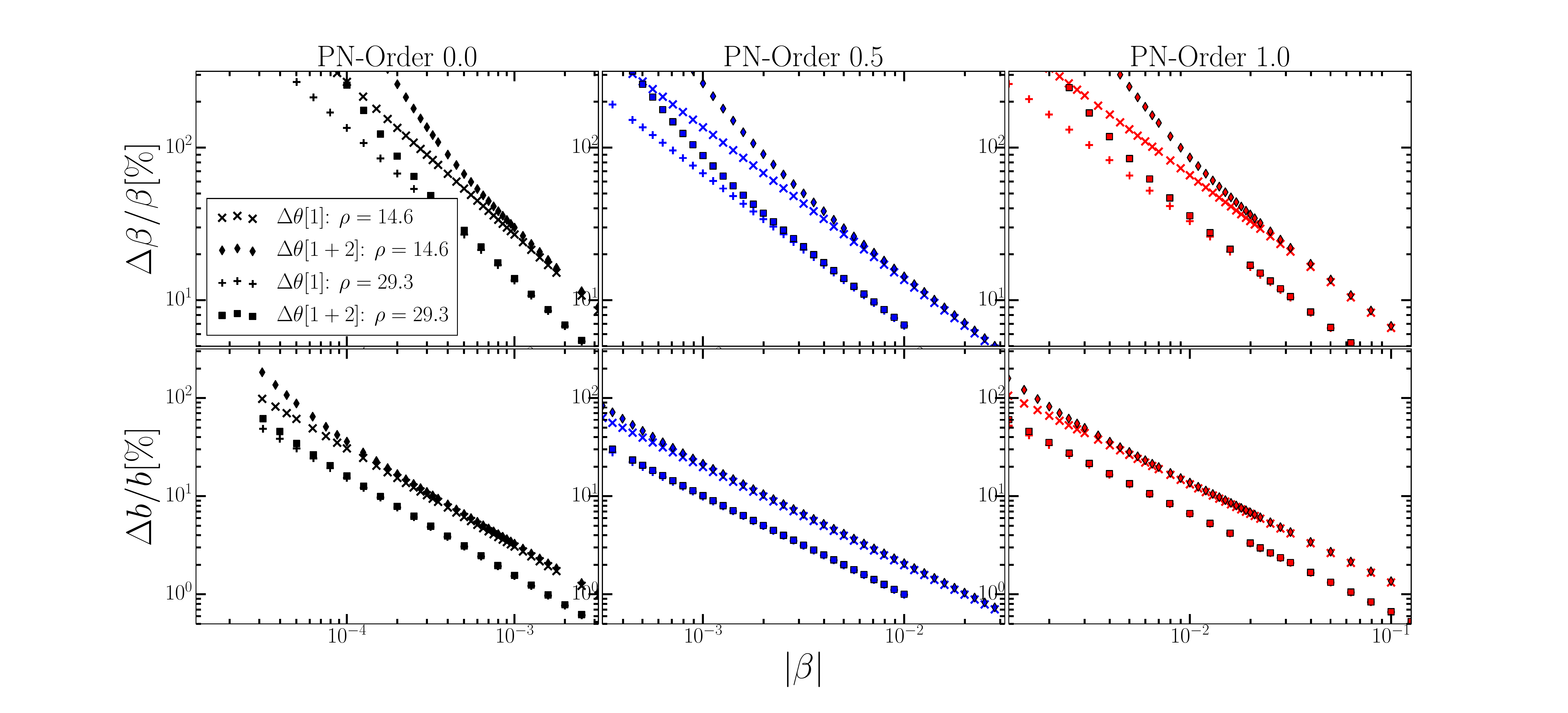}
\includegraphics[width=1.2\textwidth,center]{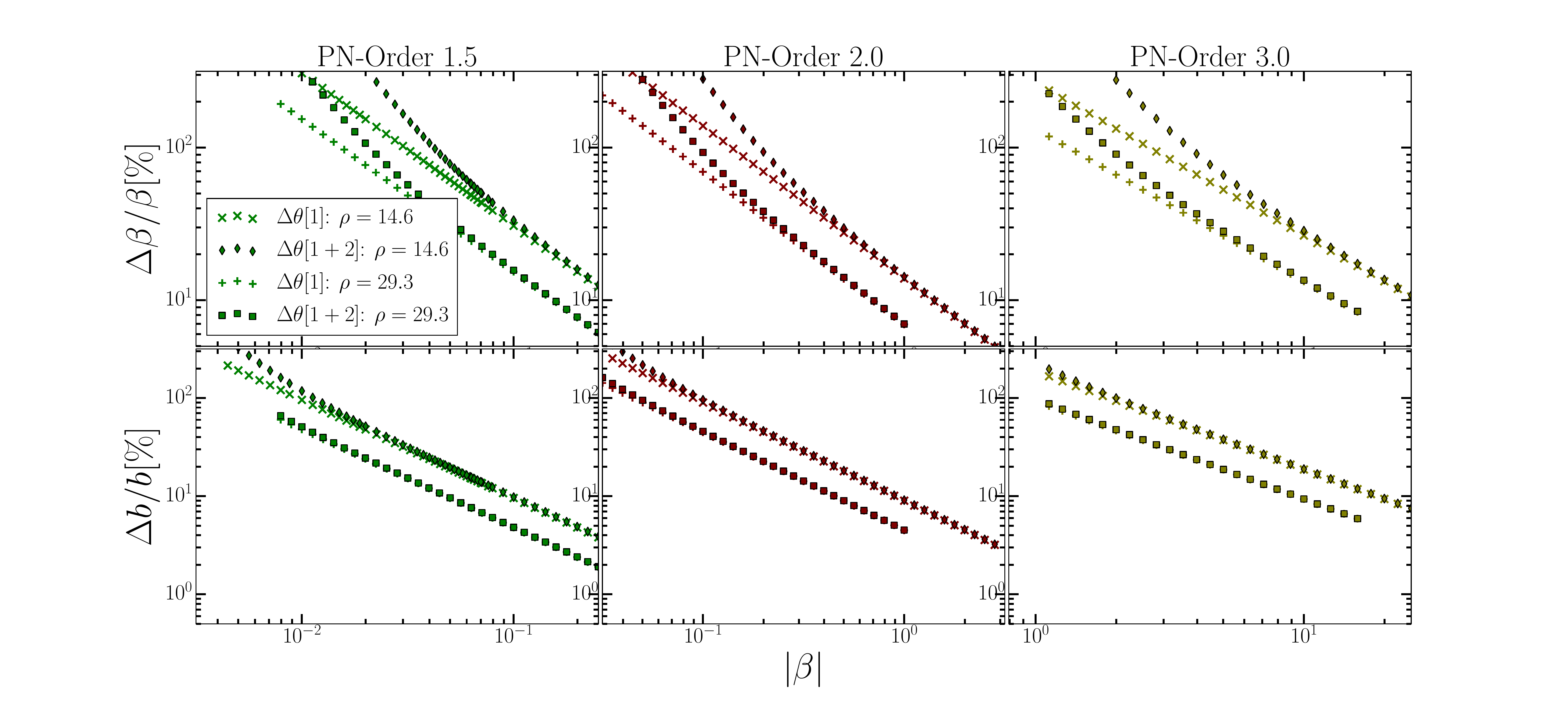}
\caption{
Sky-averaged errors as a function of $\beta$ for a two-dimensional ppE parameter space for the BBH 1:1 system of averaged network SNR $\rho = 14.6$.  SNR results of $\rho = 29.3$ are also showed by setting the distance to $D_L = 550$ Mpc.  As noted in Ref.~\cite{Vit_Zan2011_Net} error estimates are rescaled as $\sigma [1] (\rho^* / \rho) $ and $\sigma [2] (\rho^* / \rho)^2$, where $\rho^*$ is the SNR that error estimates are originally calculated from.  In the top panel the far left column represents each system for a PN-order 0.0 modification ($b=-5$), the center column is a PN-order 0.5 modification ($b=-4$), and far right column is for PN-order 1.0 modifications ($b=-3$).  Similarly, the bottom panels are resulting modifications at PN-order 1.5  ($b=-2$), 2.0  ($b=-1$), and  3.0  ($b=+1$).  $\beta$ is more tightly constrained at lower PN-orders and the inclusion of second-order errors for $(\beta,b)$ drastically diverge from Fisher estimates  as $\beta \rightarrow 0$.  
}
\label{2dimension}
\end{figure*}

\begin{figure*}
\includegraphics[width=1.1\textwidth,center]{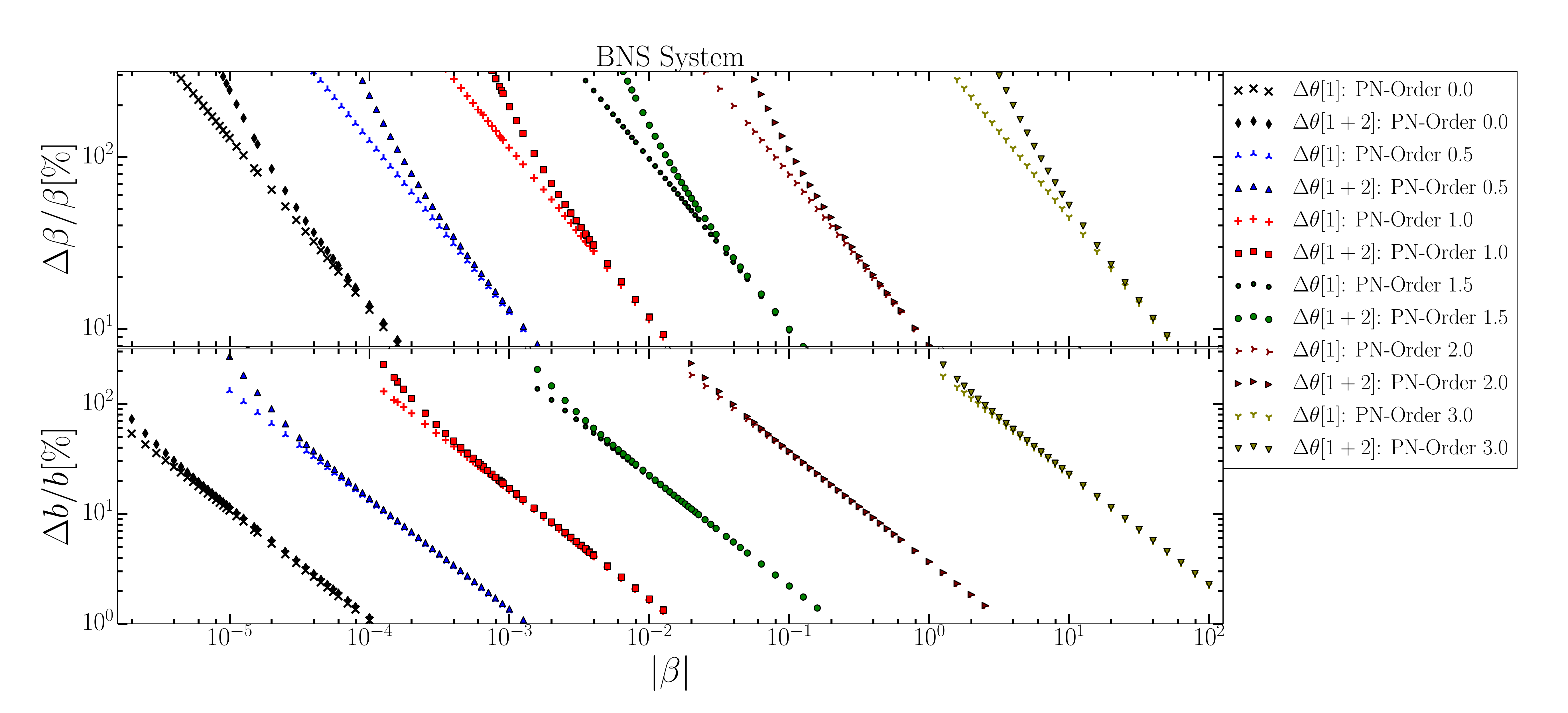}
\caption{
Sky-averaged errors, similar to figure~\ref{2dimension}, for a BNS system of averaged SNR $\rho = 17.0$.  
}
\label{2dimension_BNS}
\end{figure*}

 \begin{table*}
\begin{ruledtabular}
\begin{tabular}{c|c|c|c|c|c|c}
\;Error Bounds (System)\;	&	\;PN-order 0.0\;		&	\;PN-order 0.5 \;	&	\;PN-order 1.0 \;	&	\;PN-order 1.5\;		&	\;PN-order 2.0 \;	&	\;PN-order 3.0 \;	\\ \hline 
\;$\Delta \beta[1]$ (BBH 1:1)\;	&	$2.70 \times 10^{-4}$&	$1.36 \times 10^{-3}$&	$6.59 \times 10^{-3}$&	$3.07\times 10^{-2}$	&	$1.39\times 10^{-1}$	&	$2.66$			\\ \hline
\;$\Delta \beta[1]$ (BNS)\;		&	$1.29 \times 10^{-5}$&	$1.24 \times 10^{-4}$&	$1.14 \times 10^{-3}$&	$9.78 \times 10^{-3}$&	$7.93\times 10^{-2}$	&	$4.49$		\\
\end{tabular}
\caption{
Constant slopes of first-order error bound estimates of the BBH 1:1 (for SNR $\rho = 14.6$) and BHNS systems for all $\beta$ values.  Here percent errors [\%] follow a $1/\beta$ relationship for $\Delta \beta[1]$ represented above for respective PN-orders.
}
\label{2d_slopes}
\end{ruledtabular}
\end{table*}

\section{Results}\label{results}

In this section we explore the error bounds both as a function of the SNR and sky location of the source.  The asymptotic expansion approach is first applied to a two-dimensional ppE parameter space (when the physical parameters are known) of equal-mass systems.   Only phase corrections are assumed through unknown ppE parameters $(\beta, b)$, while $b$ probes modifications at PN-orders 0.0-3.0 of the {\texttt{TaylorF2}} model (of a PN-order 3.5 phase).  Based on Ref.~\cite{Vit_Zan_Mak_2010_single, Vit_Zan2011_Net, Vit_Zan2010_BeyondFisher} this approach is expected to give  overly optimistic errors. The Fisher information error estimates presented here  for the ppE parameters are at least an order of magnitude smaller than results with Bayesian model selection~\cite{Corn_Samp_Yun_Pret2011_GWtests}.   

 To identify SNR dependencies and regions of lowest error estimates the sky dependencies of errors are observed through a  289-point sky grid.  A point $(\mbox{lat}_{i},\mbox{long}_{j})$ in latitude-longitude coordinates (of the Earth frame) on the sky grid follows from the procedure of Ref.~\cite{Vit_Zan2011_Net} (detector coordinates also follow Ref.~\cite{Vit_Zan2011_Net}, which are fixed in the Earth Frame as given in Ref.~\cite{Allen_1996, Schutz_2011}).

As discussed in Section~\ref{param_space_expan}, $\epsilon = \pi /6$ is a fixed value and excluded in error analysis.  Parameter $\psi$ is also fixed and arbitrary values can be chosen for fiducial parameters $\phi_c$ and $t_c$.  The sky-averaged SNR is restricted to an inspiral phase $\rho <20$ to focus on the more likely advanced interferometer scenarios.  For each system considered, the distance of the resolved signal in the network is varied to keep a fixed SNR.  For a three-detector network ($I = H, L ,V$) the following is chosen for the equal-mass binary systems:
\begin{enumerate}[label=$\cdot$]
\item BBH 1:1- $(m_1 , m_2)=(10, 10) M_{\odot}, \,D_L = 1100 \mbox{Mpc}, $ 
\item BNS- $(m_1 , m_2)=(1.4, 1.4) M_{\odot}, \, D_L = 200 \mbox{Mpc}. $ 
\end{enumerate}
Here the constructed binary black hole (BBH) and binary neutron star (BNS) system leaves the network with an averaged SNR of $\rho = 14.6$ and $\rho = 17.0$, respectively.  For  unequal mass systems we choose a BBH system with a 1:2 mass ratio and a black hole neutron star (BHNS) binary with the following:
\begin{enumerate}[label=$\cdot$]
\item  BBH 1:2- $(m_1 , m_2)=(5, 10) M_{\odot}, \, D_L = 850 \mbox{Mpc}. $ 
\item  BHNS- $(m_1 , m_2)=(1.4, 10) M_{\odot}, \, D_L = 450 \mbox{Mpc}. $ 
\end{enumerate}
which respectively give SNRs of $\rho = 14.9$ and $\rho= 15.8$.   For direction reconstruction and related extrinsic parameters the network geometry is important; however, for intrinsic parameters (as with the ppE parameters) SNR gains and losses have a larger impact~\cite{Vit_Zan2011_Net}.

In the seven-dimensional study, $\beta$ is varied along $b = -3, -7$ for a BBH 1:1, 1:2, and BHNS systems.  The reason for $b=-3$ is that it simulates modifications to the dispersion of a GW (e.g., massive gravitons or  Lorentz violations~\cite{Will_1998_MG, Mirs-Yunes-Will_2012_LV}).  Also, $b=-7$ simulates weak-field modifications for dipole gravitational radiation (e.g., Brans-Dicke~\cite{Yunes_Siem_2013, Will_1994_BD}) and the non-spinning, even-parity sector of quadratic modified gravity (e.g., Einstein-Dilation-Gauss-Bonnett, or EDGB, gravity~\cite{QMG}).  Distinguishability from GR is denoted as the condition that errors are smaller than the separation between parameters of the GR-limit and that of some alternative theory.

\subsection{Two-dimenstional study: equal mass}\label{bidimen}

In this subsection uncertainties for a two-dimensional parameter space are computed for both the BBH 1:1 and BNS systems, marked by $\Delta \theta^i_{\regtext{ppE}}$.  Parameter $b$ is chosen at a fixed PN-order correction with PN-order 0.0, 0.5, 1.0, 1.5, 2.0, and 3.0 (i.e., $b = -5, -4, -3, -2, -1, +1$) while $\beta$ is varied at each PN order.  Here $\beta$ probes values small enough to induce a sky-averaged error larger than 100\% in $b$ and large enough for $\lesssim 10\%$ sky-averaged error in $\beta$.  Errors for the BBH 1:1 system are depicted in figure~\ref{2dimension}, each labeled column representing a particular PN-order modification.  Furthermore, to demonstrate the SNR dependence the BBH 1:1 system contains values for the scenario in which the SNR is doubled, for this the distance is decreased to $D_L =  550$ Mpc.  Figure~\ref{2dimension_BNS} illustrates similar results for the BNS system.  

The constant slopes of errors at first-order are catalogued in Table~\ref{2d_slopes} for each PN-order.  The computed first-order errors are consistent with statements of Ref.~\cite{Corn_Samp_Yun_Pret2011_GWtests} which demonstrate that different PN-order corrections lead to different feasible constraints on $\beta$-values. BNS systems offer tighter constraints on $\beta$ at each chosen $b$.  It is interesting to observe that scaling parameters controlling propagating modifications, e.g.~the graviton wavelength $\beta_{\regtext{MG}} \propto \lambda^{-2}_g$, are not more tightly constrained with BNS systems at shorter distances than BBH systems at larger distances.  Rather, parameters like $\beta_{\regtext{MG}}$, also depend on a distance measure and masses of the compact objects that adversely affect constraints at shorter distances and smaller masses.

\begin{figure*}
\begin{centering}
\includegraphics[width=1.15\textwidth,center]{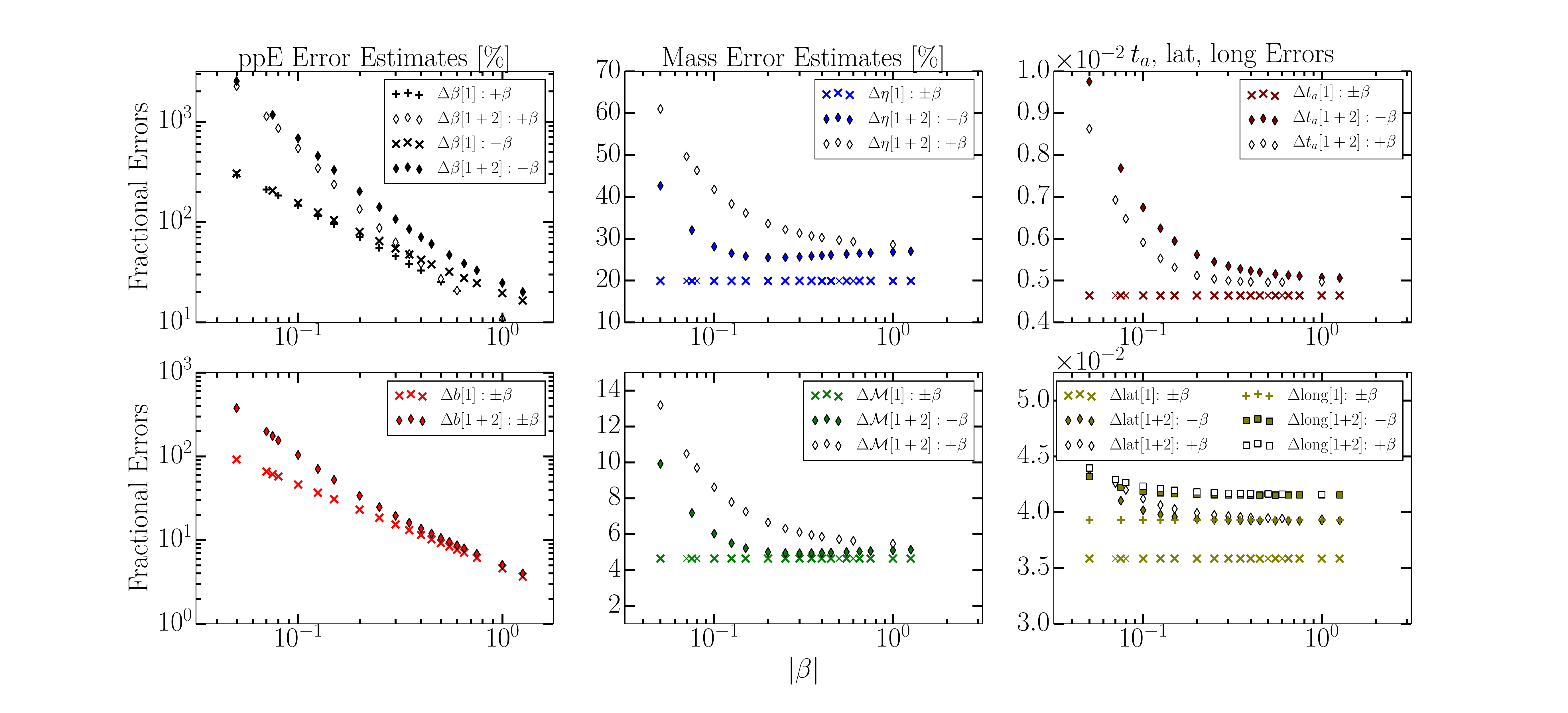}
\end{centering}
\vspace{-15pt}
\caption{
Sky-averaged uncertainties for the equal-mass BBH 1:1 system for a PN-order 1.0 modification of the seven dimensional parameter space (ppE parameters $\{ \beta, b \}$ and physical parameters $\{ \eta, \log \mathcal{M}, t_c, \mbox{lat}, \mbox{long}  \}$).  In the left column the top panel displays  $\Delta \beta$ percent errors as a function of $\beta$ (the sign of $\beta$ provides different error estimates) and below that are  $\Delta b$ errors as a function of $\beta$ (the sign of $\beta$ does not play a role in these error estimates).  In the middle and to the right are the physical parameters' errors, where the constraint of $\beta$ primarily affects the second-order contributions.   Enlarging the parameter space  increases error estimates from those computed in figure~\ref{2dimension} at PN-order 1.0, thus weakening constraints on $\beta$.  For negative $\beta$, the full-dimensional study states $\Delta \beta[1] = 100\%$ at $\beta = -0.16$ and $\Delta \beta[1+2]=100\%$ at $\beta = -0.32$.
}
\label{7dimension}
\end{figure*}

The smaller $\beta$, the more second-order effects in the errors contribute.  Second-order effects on the errors of $b$ are less significant, and only errors $>100\%$ on $\beta$ force sizeable second-order contributions in $b$.  If $b$ is near distinguishable, $\Delta b[1+2] \lesssim 100\%$, $\Delta \beta [1+2]$ are much larger than $\Delta \beta [1]$.  Only when $\Delta b[1+2] \lesssim 10\%$ do $\Delta \beta [1]$ and $\Delta \beta [1+2]$ converge to similar estimates.  Simulations producing the results of figures~\ref{2dimension} and~\ref{2dimension_BNS} used both $\pm \beta$ values and the skewed representation of figure~\ref{fitting_factor} is not apparent.  Note that the range of $\beta$ values, in which error bounds are $\leq100\%$ (figures~\ref{2dimension} and~\ref{2dimension_BNS}), are orders of magnitude smaller than the $\beta$-value ranges considered in previous studies based on Bayesian methods~\cite{Corn_Samp_Yun_Pret2011_GWtests}.

\begin{figure*}[t]
\begin{centering}
\includegraphics[scale=.23]{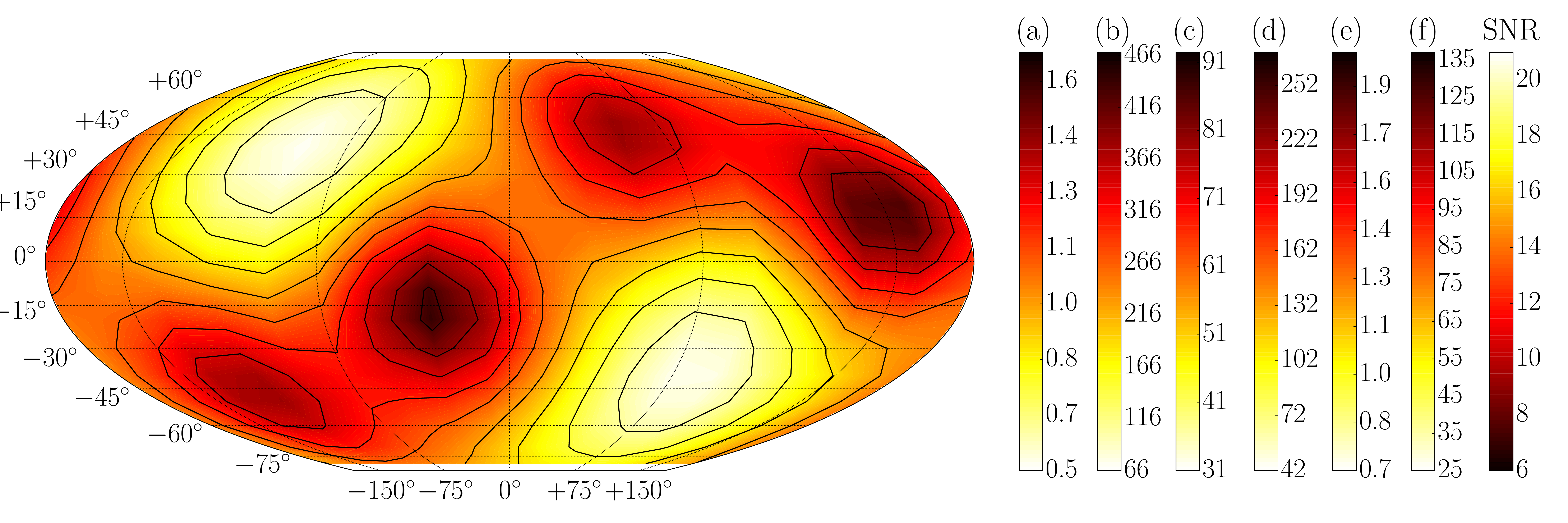}
\end{centering}
\caption{Sky distribution of error estimates.  Color bars represent range of ppE quantities labeled (a), (b)\dots, (f)  in Table~\ref{min_max_sky}.  This demonstrates the correlation of the SNR and ppE error estimation over the sky.  See text for discussion.
}
\label{skyview_beta}
\end{figure*}

\subsection{Full parameter space: equal mass}\label{7dimen}

The most realistic results come from the study of the largest resolvable parameter space. In this subsection, first- and second-order uncertainties $\Delta \vartheta^i$ of a full 7-dimensional parameter space are calculated for the equal-mass BBH 1:1 system, where $ \vartheta = \{ \theta_{\regtext{ppE}} , \theta_{\regtext{phys}} \}$.   Here $b$ is fixed to induce a PN-order 1.0 modification ($b=-3$).  Such corrections simulate effects produced by modifying  the GW dispersion relation~\cite{Yunes_Siem_2013, Mirs-Yunes-Will_2012_LV}.  Unlike the two-dimensional cases, the errors (first- and second-order) are effected by the sign of $\beta$, where sky-averaged errors for the ppE parameter pair $(\beta, b)$ are displayed in the left column of figure~\ref{7dimension}.  Errors of physical parameters affected by varying $\beta$ are depicted in the middle and right column of figure~\ref{7dimension}.   The skewed behavior of $\pm \beta$ results are representative of fitting factor results of figure~\ref{fitting_factor}.

For $\beta$ the first-order errors are not at a constant slope.   $\Delta \beta[1]$ approximately follows linear relationship: $\Delta \beta[1] \approx 0.046|\beta| + 0.15$, for negative $\beta$.  Here a $100 \%$ threshold error occurs at $\beta = -0.16$, for $\Delta \beta[1]$, and $\beta = -0.32$, for $\Delta \beta[1+2]$.  In this more realistic scenario, it can be seen that for extremely small $\beta$ values $b$ falls within its own uncertainty.  Yet, analogous to the two parameter space, a $100 \%$ error in $\Delta b[1+2]$ requires large errors in $\Delta \beta[1+2]$.  Furthermore, error estimates are at least an order of magnitude larger.  Another aspect of considering a full-dimensional parameter space are the additional error trends imparted on physical parameters (masses, arrival time, etc) when $\beta$ is varied, see the middle and right column of figure~\ref{7dimension}.

\begin{table}[b]
\begin{ruledtabular}
\begin{tabular}{c|c|c|c}
\;ppE $\beta$-value \;		&	\;Error Estimations\;			&	\;$\rho_{\regtext{max}}=20.8$ \;	&	\;$\rho_{\regtext{min}}=7.0$ \;	\\  \hline
$-0.25$					&							&							&							\\ \hline
(a)						&	$\Delta b[2] / \Delta b [1]$		&	$\;0.55\;$					&	$\;1.67\;$					\\ 
						&	$\Delta b[1]$				&	$\;12.1\;$ [\%]				&	$\;36.2\;$ [\%]				\\ 
						&	$\Delta b[1+2]$				&	$\;13.8\;$ [\%]				&	$\;70.5\;$ [\%]				\\ 
						&	$\Delta\beta[2] / \Delta \beta[1]$&	$\;1.19\;$					&	$\;3.57\;$					\\ 
						&	$\Delta\beta[1]$			&	$\;42.7\;$ [\%]				&	$\;126.4\;$ [\%]				\\ 
(b)						&	$\Delta\beta[1+2]$			&	$\;66.4\;$ [\%]				&	$\;468.7\;$ [\%]				\\ \hline
$-0.35$					&							&							&							\\ \hline
						&	$\Delta b[2] / \Delta b [1]$		&	$\;0.43\;$					&	$\;1.28\;$					\\ 
						&	$\Delta b[1]$				&	$\;8.7\;\;$ [\%]				&	$\;25.8\;$ [\%]				\\ 
						&	$\Delta b[1+2]$				&	$\;9.4\;\;$ [\%]				&	$\;42.0\;$ [\%]				\\ 
						&	$\Delta\beta[2] / \Delta \beta[1]$&	$\;0.91\;$					&	$\;2.72\;$					\\ 
(c)						&	$\Delta\beta[1]$			&	$\;31.4\;$ [\%]				&	$\;92.9\;$ [\%]				\\ 
(d)						&	$\Delta\beta[1+2]$			&	$\;42.4\;$ [\%]				&	$\;269.1\;$ [\%]				\\ \hline
$-0.55$					&							&							&							\\ \hline
						&	$\Delta b[2] / \Delta b [1]$		&	$\;0.32\;$					&	$\;0.99\;$					\\ 
						&	$\Delta b[1]$				&	$\;5.5\;\;$ [\%]				&	$\;16.4\;$ [\%]				\\ 
						&	$\Delta b[1+2]$				&	$\;5.8\;\;$ [\%]				&	$\;23.2\;$ [\%]				\\
(e)						&	$\Delta\beta[2] / \Delta \beta[1]$&	$\;0.65\;$					&	$\;1.96\;$					\\ 
						&	$\Delta\beta[1]$			&	$\;21.1\;$ [\%]				&	$\;62.4\;$ [\%]				\\ 
(f)						&	$\Delta\beta[1+2]$			&	$\;25.2\;$ [\%]				&	$\;137.3\;$ [\%]				\\
\end{tabular}
\caption{ Maxima and minima of estimates depicted in the sky-map plot (figure~\ref{skyview_beta}) for respective $\beta$-values of figure~\ref{7dimension}.  Errors are the smallest for $\rho_{\regtext{max}}=20.8$ and largest for $\rho_{\regtext{min}}=7.0$.  Terms labeled with (a), (b)\dots, (f)  correspond to respective color bars in figure~\ref{skyview_beta}.  Values are chosen because they offer the most insight. }
\label{min_max_sky}
\end{ruledtabular}
\end{table}

The sky distributions of the errors and the SNR are shown in figure~\ref{skyview_beta}.  Table~\ref{min_max_sky} catalogs this for $-\beta =0.25, 0.35, 0.55$.   This SNR dependence is similar to intrinsic parameters for GWs. The $\beta$ values, being a PN-order 1.0 correction characterizing massive graviton dispersion tests, are chosen for the following reasons:
\begin{enumerate}
\item At $\beta = -0.25$, figure~\ref{7dimension} identifies the conditions:  $\Delta b[2] / \Delta b[1] \approx 1$ with $\Delta \beta[1]<100\%<\Delta \beta[1+2]$.  Sky averages are performed before computing the ratios.  In SNR $\gtrsim 15$, we have $\Delta b[2] / \Delta b[1] \lesssim 1$, as seen in (a).  (b) diplays $\Delta \beta [1+2]$, which ranges from $66.4\%$ to $468.7\%$.  $\Delta \beta[2]$ dominates the error budget.  
\item For $\beta = -0.35$, sky-averaged $\Delta \beta[1]< \Delta \beta[1+2] \approx 100\%$.  Although $\Delta b[2]/ \Delta b[1]>1$, in limited portions of the sky, the ratio never exceeds $1.3$ with a maximum of $\Delta b [1+2] = 42.0\%$.  There is a strong increase in $\Delta \beta[1+2]$ from $\Delta \beta[1]$ in low SNRs.  The majority of the sky is dominated by second-order terms, with $\Delta \beta[2] / \Delta \beta[1]$ ranging from  $0.91$ to $2.72$.  
\item $\beta =- 0.55$ is where we calculate sky-averaged ratio $\Delta \beta[2] / \Delta \beta[1] \approx 1$  with $\Delta \beta[1] < \Delta \beta[1+2] < 100\%$.  Here larger portion of the sky has  ratio $\Delta \beta[2] / \Delta \beta[1]<1$ as shown in (e).  A majority (but not all) of the sky-map has  total error falling below $100\%$ after the inclusion of second-orders with sky-averaged error at $\Delta \beta[1+2] \approx 47\%$.
\end{enumerate}
From the known dependence on $\rho$, quantities displayed in figure~\ref{skyview_beta} and Table~\ref{min_max_sky} can be easily re-derived for higher or lower SNRs.

\begin{figure*}
\begin{centering}
\includegraphics[width=1.15\textwidth,center]{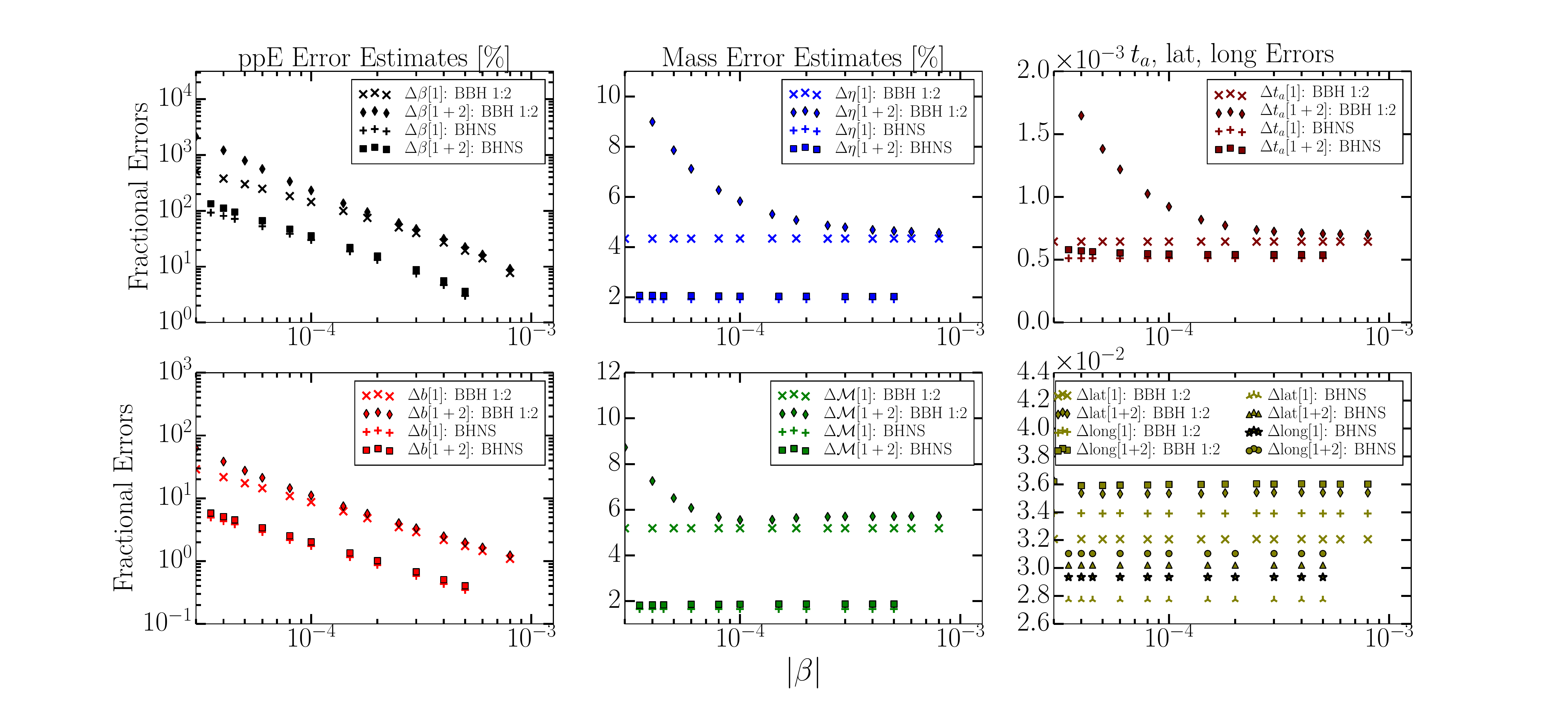}
\end{centering}
\vspace{-15pt}
\caption{
Sky-averaged error estimates for the BBH 1:2 and BHNS system.  Left column represent calculations of the ppE parameter errors ($\Delta \beta, \Delta b$) for negative $\beta$-values, center column are the mass errors ($\Delta \eta, \Delta \mathcal{M}$), and far right are arrival time $\Delta t_a$ and latitude-longitude ($\Delta \mbox{lat}, \Delta \mbox{long}$) error estimates.  Here latitude-longitude error estimates are not affected by $\beta$ variation, as was previously presented in the equal-mass system.  This study states that $\Delta \beta_{\regtext{BBH1:2} }[1+2]=95.2\%$ at  $\beta_{\regtext{BBH1:2} } = -1.8\times10^{-4}$ and $\Delta \beta_{\regtext{BHNS} }[1+2]=95.3\%$ at  $\beta_{\regtext{BHNS} } = -4.5\times10^{-5}$.
}
\label{7dimension_uneq_BBH}
\end{figure*}

\subsection{Full parameter space: unequal mass}\label{7dimen2}

Here first- and second-order uncertainties $\Delta \vartheta^i$ of a full seven-dimensional parameter space are calculated for the BBH 1:2 and BHNS system.  In this case a weak-field $b = -7$ modification is induced, which in our context mimics the non-spinning, even-parity sector of quadratic modified gravity (QMG) and can include specifics like EDGB gravity.  Inclusion of QMG modifications is due to $\beta$ being resolvable by a non-zero mass differences at this PN-order.  These modifications manifest through modification of the energy flux as $\beta \propto \zeta_3 (1-4 \eta)$~\cite{QMG} and the BHNS binary can also test examples of dipole gravitational radiation, like Brans-Dicke (BD).   

Error bounds  are presented in figure~\ref{7dimension_uneq_BBH}.   The overall trend of this system's estimates are similar to the results of the equal-mass BBH 1:1 of the previous subsection, with a few exceptions.  The first being that the separation between errors $\Delta \beta[1], \Delta b[1]$ and $\Delta \beta[1+2], \Delta b[1+2]$ are not as great as with the PN-order 1.0 modification.  In comparison to the previous subsection, the chirp mass errors $\Delta \mathcal{M}$ are roughly the same, yet $\Delta \eta$ estimates are considerably less. Time of arrival errors $\Delta t_a$ are also less and latitude-longitudinal estimates don't suffer from varying $\beta$ at first- and second-order.

\begin{figure*}[t]
\begin{centering}
\includegraphics[scale=0.4]{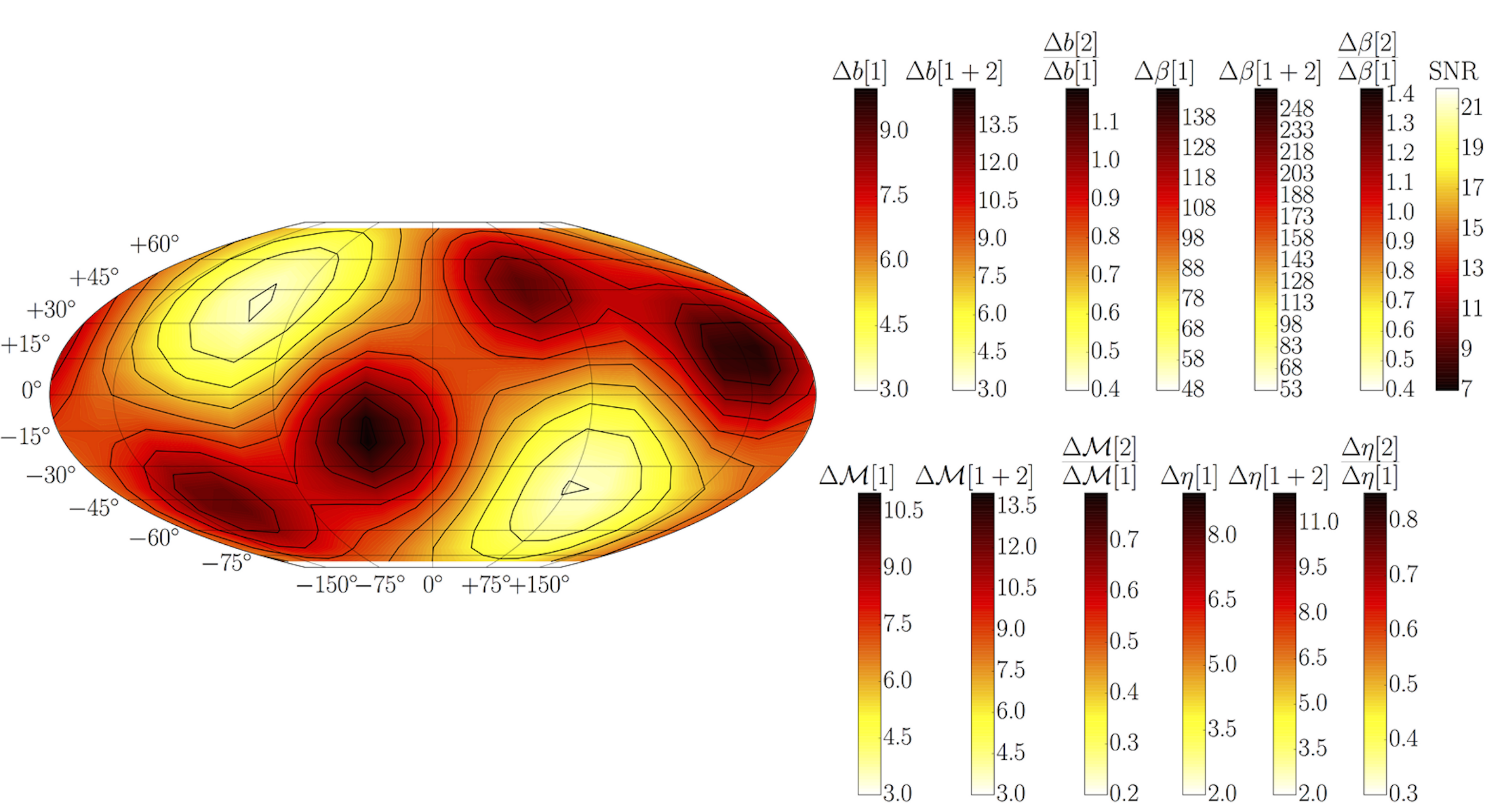}
\end{centering}
\caption{ 
Sky-map error estimates of ppE parameters $\Delta \beta$ and $\Delta b$ and mass parameters $\Delta \eta$ and $\Delta \mathcal{M}$ for the unequal mass BBH 1:2 system.  The top color bars for ppE parameters are for  $\beta_{\regtext{BBH1:2} }  = -1.8 \times 10^{-4}$ and the mass parameters below that are for $\beta_{\regtext{BBH1:2} }  = -3.0 \times 10^{-4}$ of results in figure~\ref{7dimension_uneq_BBH}. The SNR color bar is valid for both error estimates.  Sky-average estimates provide $\Delta \beta_{\regtext{BBH1:2} } [1+2]=95.2\%$, of $\beta_{\regtext{BBH1:2} }  = -1.8 \times 10^{-4}$, and $\Delta \beta_{\regtext{BBH1:2} } [1+2]=47.4\%$ at $\beta_{\regtext{BBH1:2} }  = -3.0 \times 10^{-4}$.}
\label{sky_view_uneqBBH}
\end{figure*}

For the BBH 1:2 system sky contours of ppE and mass error estimates at, respectively, $|\beta| = 1.8 \times 10^{-4}$ and $|\beta| = 3.0 \times 10^{-4}$ are displayed in figures~\ref{sky_view_uneqBBH}.  In figure~\ref{sky_view_uneqBBH}, the mass error estimates (bottom color bars) are plotted since this $\beta$-value produces sky-averaged estimate $\Delta \beta[1+2] < 100\%$, with second-order effects in the mass estimates making notable contributions (see figure~\ref{7dimension_uneq_BBH}).   We observe that in such a context second-order effects do not dominate the error budget of $\Delta \eta$ and $\Delta \mathcal{M}$ in this sky-grid.  In low-SNR regions, $\Delta \eta[2]/ \Delta \eta[1]$ and $\Delta \mathcal{M}[2]/\Delta \mathcal{M}[1]$ are near unity.  In these same low-SNR regimes $\Delta \beta[2]/ \Delta \beta[1] > 1$ and $\Delta \beta[1+2] > 100\%$, which demonstrates the sky-grid SNR relation to errors accrued on physical parameters due to large error estimates of ppE parameters. 

Figure~\ref{sky_view_uneqBBH} also represents a second set of contours generated for $|\beta| = 1.8 \times 10^{-4}$ modifications.  Top color bars are representative of ppE parameter error estimates $(\Delta \beta, \Delta b)$  valid for this choice of $\beta$.  Contours are plotted at this $\beta$-value since this simulates the condition that $\Delta \beta[1+2] \approx 100\%$ with $\Delta \beta[1] < 100\%$.  Again we observe the volatility in $\Delta \beta[1+2]$ estimates, ranging from $53\%$ to about $250\%$ while remaining strongly correlated to the SNR.  One notable feature of this plot is that ratios $\Delta b[2]/\Delta b[1]$ and $\Delta \beta[2]/\Delta \beta[1]$ are relatively close to each other, being approximately equal to each other in regions of high-SNR.  This is in contrast to the equal-mass study of the previous subsection and demonstrates the small separation in $\Delta \beta[1]$ and $\Delta \beta[1+2]$ estimates depicted in the left column of figure~\ref{7dimension_uneq_BBH}, which allows the ratio $\Delta b[2]/\Delta b[1]$ to be comparable to  $\Delta \beta[2]/\Delta \beta[1]$.  Relations between these quantities depicted in figure~\ref{sky_view_uneqBBH} can be compared to the extrema of the equal-mass BBH system of PN-order 1.0 modifications catalogued in Table~\ref{min_max_sky}.  Similar results come from the BHNS system.

\begin{figure*}
\begin{centering}
\includegraphics[width=1.15\textwidth,center]{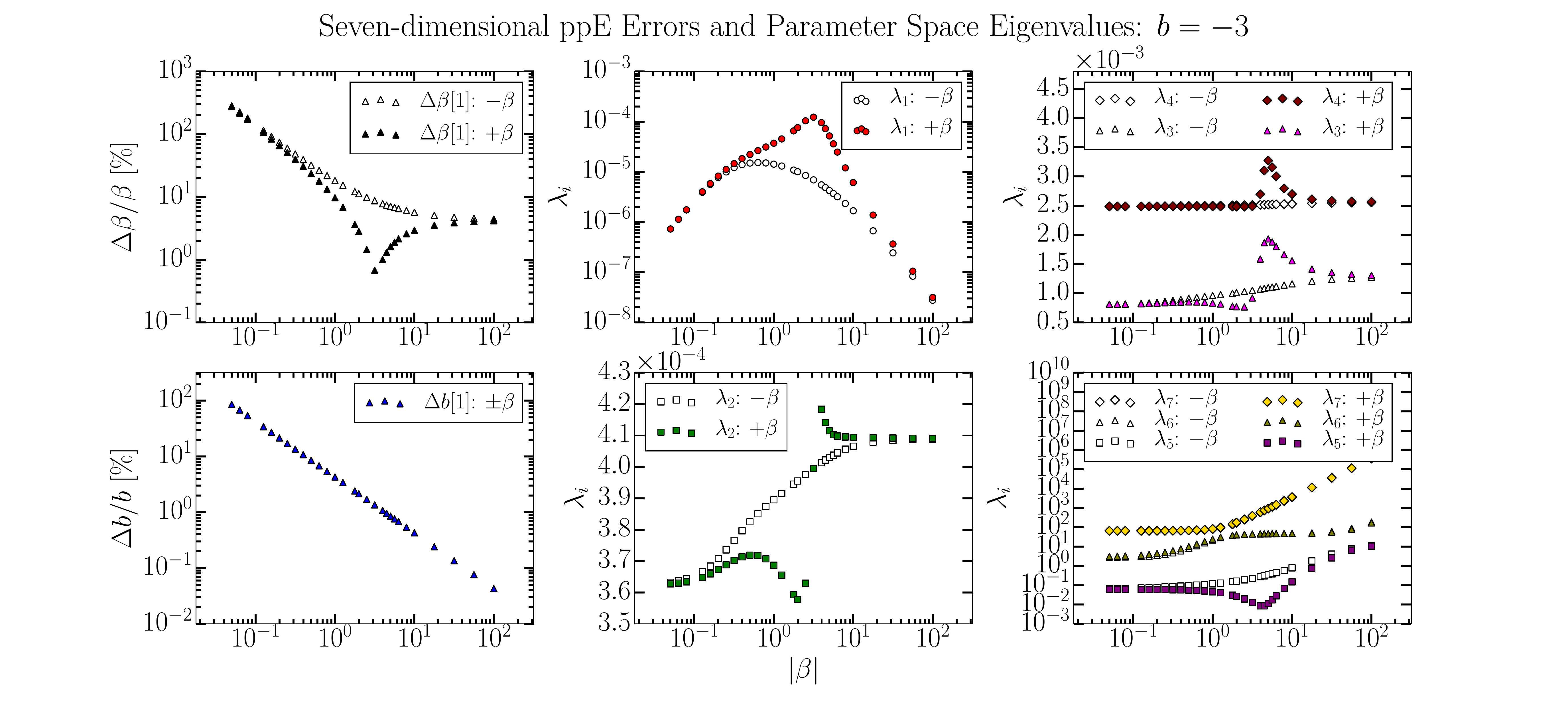}
\includegraphics[width=1.15\textwidth,center]{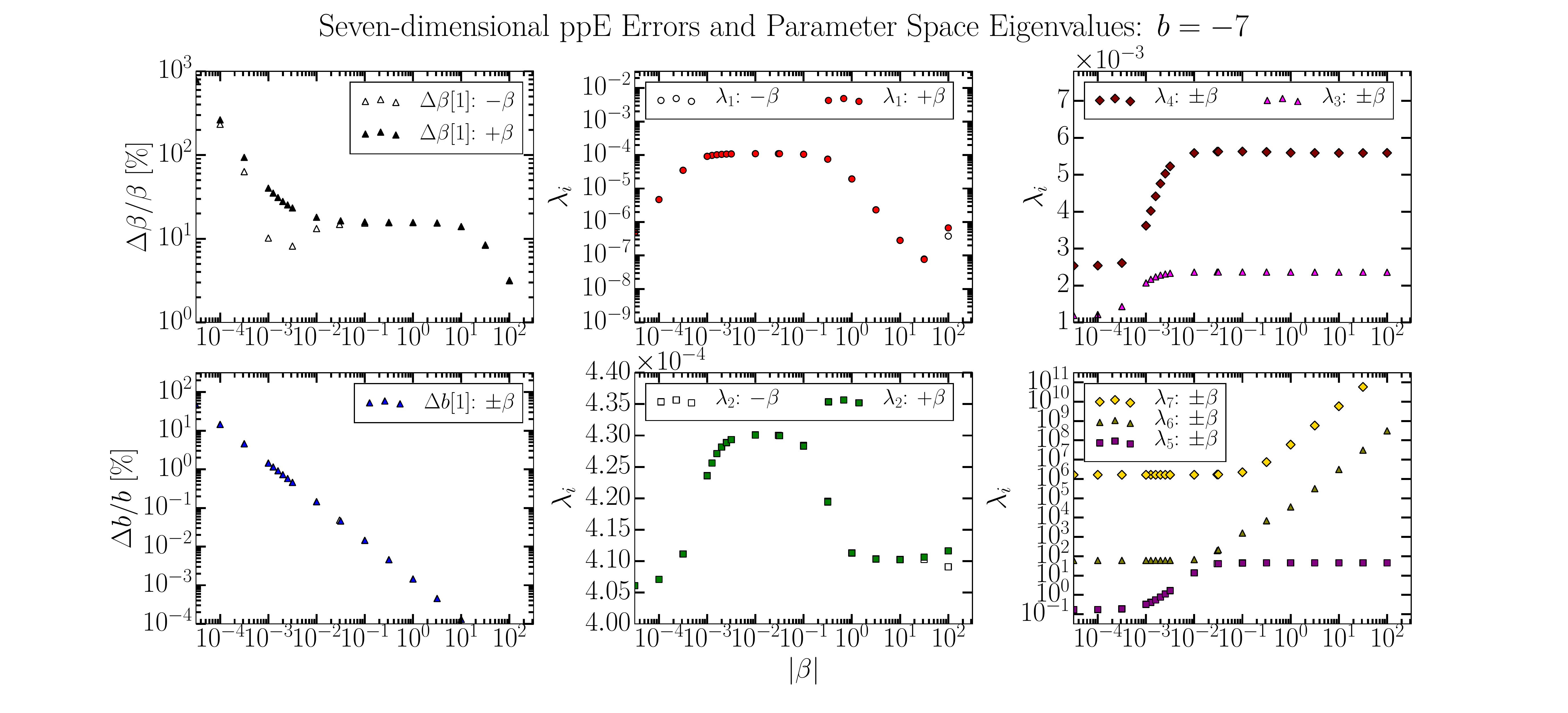}
\end{centering}
\vspace{-15pt}
\caption{
First order errors (left panels) and eigenvalues (center and right panels) of the Fisher matrix  when computations are extended to the seven dimensional parameter space.}
\label{evalues}
\end{figure*}

In order to check that the Fisher information matrix did not become singular we systematically explored its eigenvalues.  For example figure~\ref{evalues} shows scenarios in which the Fisher matrix becomes singular for the seven dimensional study.  These values of $\beta$ were avoided in this analysis.

\subsection{Application to explicit alternative theories}\label{mod_grav}

\begin{table}[b]
\begin{ruledtabular}
\begin{tabular} {  lc  }
\multicolumn{2}{ c }{ \;Distinguishability constraint ($\lesssim 100\%$ Error)\;  }			\\ \hline		
\phantom{aaaaaaaaaaa}	$\lambda_{g,\regtext{LV}} > 3.04 \times 10^{12}$ km	&	(BBH 1:1)	\\ \hline
\phantom{aaaaaaaaaaaa}	$\xi_3^{1/4} < 7.17 $ km 						&	(BBH 1:2)	\\ \hline
\phantom{aaaaaa}\,	$|\alpha_{\regtext{EDGB}}|^{1/2} < 2.69$ km 		&	(BBH 1:2)	\\  \hline
\phantom{aaaaaaaaaaaa}	$\xi_3^{1/4} < $ 9.45 km  						&	(BHNS)	\\ \hline
\phantom{aaaaaa}\,	$|\alpha_{\regtext{EDGB}}|^{1/2} < 3.55$ km		&	(BHNS)	\\ \hline
\phantom{aaaaaaaaaaaa}	$\omega_{\regtext{BD}} > 12.7  (s_{\regtext{NS}} - 0.5 )^2$  	&	 (BHNS)	\\
\end{tabular}
\caption{
Seven-dimensional study of the BBH 1:1, 1:2, and BHNS systems with feasible constraints, i.e., computed MSE $\lesssim 100\%$.   The first considers PN-order 1.0 modifications and the latter two consider $b=-7$ modifications.  Included are the graviton wavelength (or generic Lorentz-violating) dispersion modification and  non spinning, even-parity sector models of QMG (EDGB parameter included).  Brans-Dicke constraint depends on sensitivity parameter $0.2 \leq s_{\regtext{NS}} \leq 0.3$.
}
\label{bounds}
\end{ruledtabular}
\end{table}

Since the modification considered in  subsection~\ref{7dimen} occur at PN-order 1.0 in the phase, an analysis can be done from these results for the massive graviton model.  Progression of sky-averaged errors for $\Delta \beta[1+2]$, calculated from negative $\beta$-values, of figure~\ref{7dimension} imposes a constraint of $|\beta_{\regtext{MG}} | \leq 0.31$.  Existing constraints are $|\beta_{\regtext{MG, static}} | \leq 0.37$ and $|\beta_{\regtext{MG, GW}} | \leq 2.89\times10^{-2}$, based on  current static and dynamical (from GW150914 event) bounds on $\lambda_g$ (see section~\ref{ppE_model}) computed from the BBH 1:1 system at 1100 Mpc.  This asymptotic approach thus produces an additional $16.2\%$ constraint on existing static bounds at $1\sigma$.  When including second-order terms in error estimation the constraints on $\lambda_g$ have a fractional increase of $30\%$ from the first-order Fisher matrix approach as calculated in this paper.  Given these results, further constraints on the graviton wavelength $\lambda_g$ may be possible, even with second-order error terms accounted for in the low-SNR limit of the inspiral stage only.  From calculated results the sky-averaged feasible bounds are displayed in Table~\ref{bounds}. 

 Bayesian assessments in the ppE framework of unequal mass systems (of 1:2 and 1:3 ratios) with SNR of 20 put constraints at $\lambda_g > 8.8 \times 10^{12}$ km~\cite{Corn_Samp_Yun_Pret2011_GWtests}.   Other Bayesian studies also conclude that advanced detecters would generally not favor a MG theory over that of GR when $\lambda_g$  is larger than the most stringent static bounds~\cite{Pozzo_Veitch_Vecchio2011}.  From the TIGER method implemented in the testing GR analysis of GW150914, constraints are at $\lambda_g > 10^{13}$ km, when the full inspiral-merger-ringdown signal is used (total SNR of $\rho \sim 24$)~\cite{test_gr}.  In this respect, our errors impart a more conservative approach to error estimation that still suggest that constraints may still be improved.  
   
An application of seven-dimensional results presented in subsection~\ref{7dimen2} for the  BBH 1:2  can also be made.  This $b=-7$ modification has $\beta_{\regtext{QMG}} \propto \zeta_3 (1-4 \eta)$.  In this context the constraint parameter is $\zeta_3 = \xi_3 M^{-4}$ in the non-spinning, even-parity sector of QMG, where $\xi_3 = 16 \pi \alpha_{\regtext{EDGB}}^2  $ in EDGB gravity~\cite{QMG}.  For the BBH 1:2 system figure~\ref{7dimension_uneq_BBH} presents $\Delta \beta[1] = 99.7\%$ at $|\beta|= 1.4 \times 10^{-4}$ and $\Delta \beta[1+2] = 95.2\%$ at $|\beta|= 1.8 \times 10^{-4}$.  These computations translate to respective inputs in Table~\ref{bounds} for $\xi_3$ and $\alpha_{\regtext{EDGB}}$.  Strongest suggested constraints have, in terms of EDGB parameter, $|\alpha_{\regtext{EDGB}}|^{1/2} < 1.9$ km and $|\alpha_{\regtext{EDGB}}|^{1/2} < 9.8$ km~\cite{xray_binary, EDGB_bound1}.   In weak-field tests the Cassini spacecraft has provided $|\alpha_{\regtext{EDGB}}|^{1/2} < 8.9 \times 10^6$ km (i.e., $\xi_3^{1/4}<  2.4 \times 10^7$ km)~\cite{cassini}.  Bayesian results estimate $\xi_3^{1/4} \lesssim 11$ km (or $|\alpha_{\regtext{EDGB}}|^{1/2} \lesssim 4$ km) at an SNR of 20~\cite{QMG} which is quoted in Ref.~\cite{Yunes_Siem_2013} as $\xi_3^{1/4} \lesssim 20$ km for an SNR of 10.  

Similar application to QMG and EDGB theories can be done with results of the BHNS system.  These constraints are also presented in Table~\ref{bounds} and are more stringent than the BBH 1:2 system.   With BHNS systems Brans-Dicke can be investigated through $\beta_{\regtext{BD}} \propto (s_1 - s_2)^2 \omega_{\regtext{BD}}^{-1}$, where constraint parameter is $\omega_{\regtext{BD}}$ with $s_{\regtext{BH}} = 0.5$ for black holes and for neutron stars $0.2 \leq s_{\regtext{NS}} \leq 0.3$~\cite{Healy_2012, Mirshekari_Will_2013, Yunes_Pani_Cardoso_2012, Zaglauer_1992}.   Figure~\ref{7dimension_uneq_BBH} results indicate $\Delta \beta[1] = 95.3\%$ at $|\beta|= 4.5 \times 10^{-5}$ for the BHNS system.  Thus, constraints results in $\omega_{\regtext{BD}} \geq 1.14$ and $\omega_{\regtext{BD}} \geq 0.51$ at $s_{\regtext{NS}} = 0.2$ and $s_{\regtext{NS}} = 0.3$, respecitvely.  Results of the Cassini spacecraft have also established $\omega_{\regtext{BD}} > 4 \times 10^{4}$~\cite{cassini2}.  In Ref.~\cite{Will_1994_BD} Fisher estimates placed constants of $\omega_{\regtext{BD}} > 194$ for a BHNS systems of similar masses.  

\section{Conclusion}\label{conclusion}

\begin{figure}
 \begin{centering}
\includegraphics[scale=0.29]{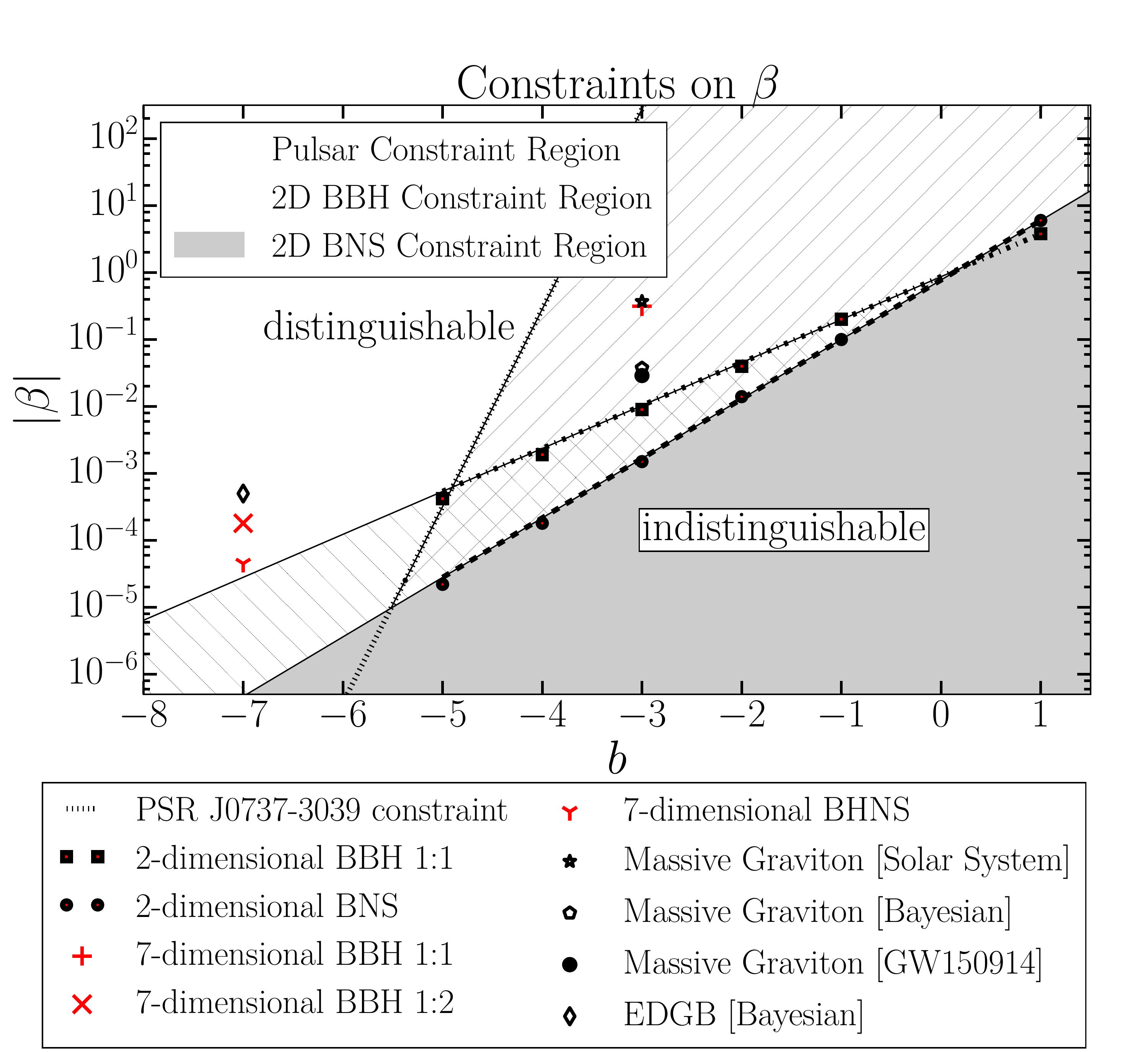}
\end{centering}
\vspace{-.1in}
\caption{
Constraints on ppE parameters $(\beta, b)$.  Alongside frequentist mean-squared error $\lesssim 100\%$ estimates are constraints imposed by Bayesian estimates~\cite{Corn_Samp_Yun_Pret2011_GWtests}, solar system tests~\cite{Bert-Gair-Ses_2011_MG}, binary pulsar measurements~\cite{binary_pulsar_ppE, binary_pulsar}, and GW150914 event~\cite{test_gr}.  Regions below each mark/line are where violations cannot be detected based on each respective study.  The GR-limit is $\beta = 0$.  Our frequentist two-dimensional study considers ppE parameter space $(\beta, b)$, while seven-dimensional studies includes physical parameters (masses, etc.).  See text for discussion.
}
\label{summary}
\end{figure}

In this paper we implement a frequentist asymptotic expansion method to estimate error bounds on the set of ppE parameters modifying the phase of the inspiral part of low-SNR ($\rho \sim 15 - 17$) GW transients.  Figure~\ref{summary} provides a summary of the main results of this paper.  The bound on the mean-squared error estimates from compact binaries studied is shown.  Each mark represents the boundary of the $(\beta, b)$-parameter space where the minimum mean-squared error estimates are $100\%$, with $\beta$ values below each $b$-value $>100\%$ and therefore not resolvable.  Previous Bayesian studies correspond to the range of exponential ppE parameter: $-11 \leq b \leq 2$, as compared to the figure~\ref{summary} summary.  The fact that for the massive graviton case  ($b=-3$) our approach here, which is a more realistic lower limit of the Cram\'{e}r-Rao Lower Bound for early detections, rules out results that were allowed by a Bayesian study~\cite{Corn_Samp_Yun_Pret2011_GWtests}, seems to indicate the need of a careful evaluation of the role of the priors.

Results of the higher order asymptotic analysis of the frequentist approach to error estimation states that further constraints can be imposed on existing non-GR theories with the study of the seven-dimensional parameter space (see Table~\ref{bounds}).  This approach does not involve the use of priors.  Here the graviton wavelength can be constrained by an additional $16.2\%$ as compared to current static bounds~\cite{Bert-Gair-Ses_2011_MG}.  Yet, these projected constraints do not further bound the graviton wavelength when compared to Bayesian estimates or values imposed by GW150914~\cite{test_gr}.  Note that although GW150914 provides a constraint of $\lambda_g > 10^{13}$ km, our result holds for a lower SNR of the inspiral stage only.  Further studies present the  scenario for the weak-field $b=-7$ modification, which can include quadratic modified gravity (QMG) (specifics being EDGB gravity) and Brans-Dicke type modifications (figure~\ref{7dimension_uneq_BBH}).  For the non-spinning, even-party sector of QMG,  bounds suggest further constraints are possible as compared to current bounds placed by Bayesian estimates and Cassini constraints.  Furthermore, error estimates for modifications at both PN-order 1.0 and the $b=-7$ weak-field follow similar sky-map contours, which are correlated to the SNR patterns (see figures~\ref{skyview_beta} and~\ref{sky_view_uneqBBH}). 

General results show that for successively higher PN-order modifications, set by $b$, the separation between first- and second-order errors increase (see figures~\ref{2dimension} and~\ref{2dimension_BNS}).  Such an effect percolates to the seven-dimensional study.  Error bounds also increase as the parameter space is enlarged, where the two-dimensional studies provide overly optimistic error bounds.  As constraints on $\beta$ become tighter in the seven-dimensional studies, the effects of second-order  estimates also accrue on physical parameters, namely $\eta$, $\mathcal{M}$, $t_a$, and latitude-longitude parameters (see figures~\ref{7dimension} and~\ref{7dimension_uneq_BBH}).  Finally, SNR increases translate error estimates as discussed in Ref.~\cite{Vit_Zan2011_Net} (figure~\ref{2dimension}), so all results can be rescaled as a function of the SNR.

Calculations performed in this paper are for single detection scenarios.  With multiple detections the presence of  weak, but consistent, violations could be combined to a make a stronger statement about error estimations.  Such methods to {\it resolve} consistent signals were explored in a Bayesian framework in Ref.~\cite{Poz_et_al_2012} and it is left for future studies in the frequentist framework.   Furthermore, as waveform models advance, for both the inspiral and ppE framework, the application of our maximum likelihood estimator asymptotic expansion could be applied to spinning binaries or to waveforms that include the merger and ringdown phases.  This will add insight into additional modified theories mappable into the ppE framework.  

\section{Acknowledgements}

The authors would like to thank S.\,Vitale, T.G.F.\,Li, A.J.\,Weinstein, W.D.\,Pozzo, L.\,Stein, and K.\,Yagi for useful discussion and comments.  
R.\,Tso is supported by the National Science Foundation Graduate Research Fellowship Program under Grant No.\,DGE-1144469, the Ford Foundation Predoctoral Fellowship, and the Gates Foundation.

\appendix 

\section{Notation and network signal}\label{appendA}

Masses of each compact body are labeled as $m_{1,2}$,  the total mass being $M = m_1 + m_2$ with $\nu = ( \pi M f )^{1/3}$ and $\eta = m_1 m_2/ M^2$ as the reduced mass frequency and symmetric mass ratio, respectively.  The usual chirp mass is $\mathcal{M} = \eta^{3/5} M$.  Geometrized units ($G=c=1$) are also employed~\cite{MTW}.  Terms labeled with $I$ indicate a particular quantity for that $I$-th detector, e.g., $s^I$ is a signal received at some $I$-th detector, $\rho^I$ is a detector-dependent SNR, etc.  Finally, the  detectors considered are those for Adv.\,LIGO and Adv.\,Virgo, so we have $I = H, L, V $ for the respective advanced interferometers in Hanford USA, Livingston USA, and Cascina Italy.  Quantities summed over $I$ indicate the total network contribution of that term, e.g., network SNR, network Fisher matrix.  Apart from units employed  notation follows that of Ref.~\cite{maggiore}.

To discuss some of the terms appearing in~(\ref{freq_sig_GR}): $\tau_I$ is a time lag parameter accounting for the delay in the waveform's propagation from the $I$-th detector frame (IDF) to some fiducial frame (FF),\footnote{FF is the frame in which the origins are referenced to coincide.} with  $\mu^I$ and $\Phi_0^I$ being coefficients that depend on the inclination angle $\epsilon$ of the binary system and the generalized antenna patterns $\mathcal{F}^I_{+,\times}$ of each detector.  These are represented by,
\begin{eqnarray}
\tau_I  &=&  \hat{\bf n} \cdot \left( {\bf r}_I - {\bf r}_{FF}  \right), \\
\mu^I &=& \left(  \left( \frac{1}{2} \mathcal{F}_{+}^I ( 1+\cos^2 \epsilon ) \right)^2 + \left( \mathcal{F}_{\times}^I \cos \epsilon \right)^2 \right)^{1/2}, \\
\Phi_0^I &=& \arctan \frac{2 \mathcal{F}_{\times}^I \cos \epsilon}{ \mathcal{F}_{+}^I (1+\cos^2\epsilon)},
\end{eqnarray}
with $\hat{\bf n}$ the direction of travel of the waveform, ${\bf r}_I$ the distance to the $I$-th detector (i.e., the IDF origin), and ${\bf r}_{FF}$ the distance to the FF origin.  Reasons for construction of a frame of common origin is due to the feasibility and efficiency displayed in calculations of quantities in particular frames.  Notion of a common origin between the frames is valid since approximative measures\footnote{Through reasonable assumption of zero curvature over the course of the GW's propagation and introduction of time lag $\tau_I$.} allow the origins of the coordinate systems to coincide.  With respect to Ref.~\cite{Vit_Zan2011_Net} the frames are established as the already mentioned IDF and FF, with a third frame called the wave-frame (WF).\footnote{Determined through the GW's direction of travel and orthonormal WF unit vectors along its axis, where  dominant harmonic polarizations in the waveform is assumed}  In producing calculable quantities the frames are then fixed to values of that in the Earth frame (EF).  

Since the origins of the frames coincide transformation between the frames is feasible through simple Eulerian angles with the usual ZXZ convention~\cite{goldstein_mech}.  From this, a set of Euler angles $( \phi, \theta, \psi )$ converts a quantity from the FF into the WF and another set $( \alpha^I, \beta^I, \gamma^I)$ converts from the FF into the IDF through the usual rotation matrices.  Here angle $\psi$ is the polarization angle.  A variety of relations can be uncovered after defining a few new angles.  Let angle pairs $(\Phi, \Theta)$ and $(\mbox{long},\mbox{lat})$ describe the sources location in the sky (the former being in spherical coordinates and the latter in longitude-latitude coordinates), let $(\Xi, \zeta)$ be defined from projections of $\hat{\bf n}$ onto the FF's axis, define angles $(\Omega^I,\Upsilon^I)$ so that they prescribe the location of the $I$-th detector with respect to the FF, and allow angle $\Delta^I$ to span the region between the first detector arm (in the IDF) and the local northern direction.  These relations are summarized as follows:
\begin{eqnarray}\label{euler_1}
\phi &=& \Phi - \frac{\pi}{2} = \mbox{long} - \frac{\pi}{2} = \Xi + \frac{\pi}{2}  \\
\theta &=& \pi - \Theta = \frac{\pi}{2} + \mbox{lat} = \zeta \nonumber 
\end{eqnarray}
and
\begin{equation}\label{euler_2}
\alpha^I = \Omega^I + \frac{\pi}{2}, \,\,\,\,
\beta^I = \frac{\pi}{2} - \Upsilon^I , \,\,\,\,
\gamma^I = \Delta^I + \frac{\pi}{2}.
\end{equation} 

Formulation of $\mathcal{F}^I_{+,\times}$ into a symmetric-trace-free base has been performed, with respect to the Eulerian angle dependence, and what surfaces in the frequency represented signal are the two generalized antenna patterns:
\begin{eqnarray}
\mathcal{F}^I_{+}	&=&	\frac{1}{2}\Big( T_{2s}(\alpha^I, \beta^I, \gamma^I) + T_{-2s}(\alpha^I, \beta^I, \gamma^I)  \Big) \\
				&& \times   \Big( T^*_{2s}(\phi, \theta, \psi) + T^*_{-2s}(\phi, \theta, \psi)  \Big) \nonumber \\
\mathcal{F}^I_{\times}&=&	\frac{i}{2}\Big( T_{2s}(\alpha^I, \beta^I, \gamma^I) + T_{-2s}(\alpha^I, \beta^I, \gamma^I)  \Big) \\
				&& \times   \Big( T^*_{2s}(\phi, \theta, \psi) - T^*_{-2s}(\phi, \theta, \psi)  \Big) \nonumber 
\end{eqnarray}
where $T_{mn}$ are second-order Gel'fand functions ($T^*_{mn}$ being their complex conjugates). Function statements, such as $f(\alpha^I, \beta^I, \gamma^I)$ and $g(\phi, \theta, \psi)$, represent their dependencies on Euler angle rotations from $FF \rightarrow IDF$ and $FF \rightarrow WF$, respectively.  See Ref.~\cite{Vit_Zan2011_Net} for exemplary calculations.  Note that an auxiliary ppE template has been developed that considers extra polarizations of waveforms produced in non-GR gravity, incorporating additional propagating degrees of freedom in the ppE framework~\cite{Chat_Yun_Corn2012_ExtendedppE}.  Although it is of interest to measure extra polarizations expected in a variety of alternative theories of gravity, these extra modes lead to more complex models.  For initial analysis of modified gravity through the asymptotic maximum likelihood estimator approach a ppE template, with only the standard two propagating modes, is considered both sufficient and satisfactory for now.  Ref.~\cite{CW_pol} investigated  methods to test non-GR polarizations via continuous waveforms from asymmetric pulsars.


 \bibliographystyle{apsrev}

\begin{thebibliography}{10}

\expandafter\ifx\csname natexlab\endcsname\relax\def\natexlab#1{#1}\fi
\expandafter\ifx\csname bibnamefont\endcsname\relax
  \def\bibnamefont#1{#1}\fi
\expandafter\ifx\csname bibfnamefont\endcsname\relax
  \def\bibfnamefont#1{#1}\fi
\expandafter\ifx\csname citenamefont\endcsname\relax
  \def\citenamefont#1{#1}\fi
\expandafter\ifx\csname url\endcsname\relax
  \def\url#1{\texttt{#1}}\fi
\expandafter\ifx\csname urlprefix\endcsname\relax\def\urlprefix{URL }\fi
\providecommand{\bibinfo}[2]{#2}
\providecommand{\eprint}[2][]{\url{#2}}

\bibitem[{\citenamefont{LIGO}(2010)}]{ADV_LIGO}
\bibinfo{author}{\bibnamefont{G.M.}} \bibnamefont{Hardy (for the LIGO Scientific Collaboration)},
\bibinfo{journal}{{ Classical Quantum Gravity}}
\textbf{27},~\bibinfo{journal}{084006}
(\bibinfo{year}{2010}).

\bibitem[{\citenamefont{LIGO}(2013)}]{LIGO}
\bibinfo{title}{{ https://www.advancedligo.mit.edu}}

\bibitem[{\citenamefont{VIRGO}(2013)}]{VIRGO}
\bibinfo{title}{{ https://wwwcascina.virgo.infn.it/advirgo}}

\bibitem[{\citenamefont{Abbott-et-al}(2016)}]{det_paper}
\bibinfo{author}{\bibnamefont{B.P.}} \bibnamefont{Abbott, et al},
\bibinfo{journal}{{Phys.~Rev.~Lett.}}
\textbf{116},~\bibinfo{journal}{061102}
(\bibinfo{year}{2016}).

\bibitem[{\citenamefont{Will}(2006)}]{Will_grav_LRR}
\bibinfo{author}{\bibnamefont{C.M.}} \bibnamefont{Will},
\bibinfo{journal}{{ Living Rev.\,Relativity}}
\textbf{9},~\bibinfo{journal}{3}
(\bibinfo{year}{2006}).

\bibitem[{\citenamefont{Abbott-et-al}(2016)}]{test_gr}
\bibinfo{author}{\bibnamefont{B.P.}} \bibnamefont{Abbott, et al},
\bibinfo{journal}{{ LIGO Document}}
\bibinfo{journal}{P1500213,}
\bibinfo{journal}{arXiv:1602.03841}
(\bibinfo{year}{2016}).

\bibitem[{\citenamefont{Gair-et-al}(2013)}]{Gair_et_al_LRR}
\bibinfo{author}{\bibnamefont{J.R.}} \bibnamefont{Gair},
\bibinfo{author}{\bibnamefont{M.}} \bibnamefont{Vallisneri},
\bibinfo{author}{\bibnamefont{S.L.}} \bibnamefont{Larson},
\bibinfo{author}{\bibnamefont{J.G.}} \bibnamefont{Baker},
\bibinfo{journal}{{ Living Rev.\,Relativity}}
\textbf{16},~\bibinfo{journal}{7}
(\bibinfo{year}{2013}).

\bibitem[{\citenamefont{Yunes-Siemens}(2013)}]{Yunes_Siem_2013}
\bibinfo{author}{\bibnamefont{N.}} \bibnamefont{Yunes},
\bibinfo{author}{\bibnamefont{X.}} \bibnamefont{Siemens},
\bibinfo{journal}{{ Living Rev.\,Relativity}}
\textbf{16},~\bibinfo{journal}{9}
(\bibinfo{year}{2013}).

\bibitem[{\citenamefont{Vitale-Zanolin}(2010)}]{Vit_Zan2010_BeyondFisher}
\bibinfo{author}{\bibnamefont{S.}} \bibnamefont{Vitale},
\bibinfo{author}{\bibnamefont{M.}} \bibnamefont{Zanolin},
\bibinfo{journal}{{ Phys.\,Rev.\,D}}
\textbf{82},~\bibinfo{journal}{124065}
(\bibinfo{year}{2010}).

\bibitem[{\citenamefont{Zanolin-Vitale-Makris}(2010)}]{Vit_Zan_Mak_2010_single}
\bibinfo{author}{\bibnamefont{M.}} \bibnamefont{Zanolin},
\bibinfo{author}{\bibnamefont{S.}} \bibnamefont{Vitale},
\bibinfo{author}{\bibnamefont{N.}} \bibnamefont{Makris},
\bibinfo{journal}{{ Phys.\,Rev.\,D}}
\textbf{81},~\bibinfo{journal}{124048}
(\bibinfo{year}{2010}).

\bibitem[{\citenamefont{Vitale-Zanolin}(2011)}]{Vit_Zan2011_Net}
\bibinfo{author}{\bibnamefont{S.}} \bibnamefont{Vitale},
\bibinfo{author}{\bibnamefont{M.}} \bibnamefont{Zanolin},
\bibinfo{journal}{{ Phys.\,Rev.\,D}}
\textbf{84},~\bibinfo{journal}{104020}
(\bibinfo{year}{2011}).

\bibitem[{\citenamefont{Yunes-Pretorius}(2009)}]{Yun_Pret2009_FundBias}
\bibinfo{author}{\bibnamefont{N.}} \bibnamefont{Yunes},
\bibinfo{author}{\bibnamefont{F.}} \bibnamefont{Pretorius},
\bibinfo{journal}{{ Phys.\,Rev.\,D}}
\textbf{80},~\bibinfo{journal}{122003}
(\bibinfo{year}{2009}).

\bibitem[{\citenamefont{Arun-et-al}(2006)}]{Arun_et_al_2006}
\bibinfo{author}{\bibnamefont{K.G.}} \bibnamefont{Arun},
\bibinfo{author}{\bibnamefont{B.R.}} \bibnamefont{Iyer},
\bibinfo{author}{\bibnamefont{M.S.S.}} \bibnamefont{Qusailah},
\bibinfo{author}{\bibnamefont{B.S.}} \bibnamefont{Sathyaprakash},
\bibinfo{journal}{{ Classical Quantum Gravity}}
\textbf{23},~\bibinfo{journal}{L37-L43}
(\bibinfo{year}{2006}).

\bibitem[{\citenamefont{Arun-et-al}(2006)}]{Arun_et_al_2006_2}
\bibinfo{author}{\bibnamefont{K.G.}} \bibnamefont{Arun},
\bibinfo{author}{\bibnamefont{B.R.}} \bibnamefont{Iyer},
\bibinfo{author}{\bibnamefont{M.S.S.}} \bibnamefont{Qusailah},
\bibinfo{author}{\bibnamefont{B.S.}} \bibnamefont{Sathyaprakash},
\bibinfo{journal}{{ Phys.\,Rev.\,D}}
\textbf{74},~\bibinfo{journal}{024006}
(\bibinfo{year}{2006}).

\bibitem[{\citenamefont{Arun-et-al}(2010)}]{Arun_et_al_2010}
\bibinfo{author}{\bibnamefont{C.K.}} \bibnamefont{Mishra},
\bibinfo{author}{\bibnamefont{K.G.}} \bibnamefont{Arun},
\bibinfo{author}{\bibnamefont{B.R.}} \bibnamefont{Iyer},
\bibinfo{author}{\bibnamefont{B.S.}} \bibnamefont{Sathyaprakash},
\bibinfo{journal}{{ Phys.\,Rev.\,D}}
\textbf{82},~\bibinfo{journal}{064010}
(\bibinfo{year}{2010}).



\bibitem[{\citenamefont{Pozzo-et-al}(2012)}]{Poz_et_al_2012}
\bibinfo{author}{\bibnamefont{T.G.F.}} \bibnamefont{Li},
\bibinfo{author}{\bibnamefont{W.}} \bibnamefont{Del Pozzo},
\bibinfo{author}{\bibnamefont{S.}} \bibnamefont{Vitale},
\bibinfo{author}{\bibnamefont{C.}} \bibnamefont{VanDenBroeck},
\bibinfo{author}{\bibnamefont{M.}} \bibnamefont{Agathos},
\bibinfo{author}{\bibnamefont{J.}} \bibnamefont{Veitch},
\bibinfo{author}{\bibnamefont{K.}} \bibnamefont{Grover},
\bibinfo{author}{\bibnamefont{T.}} \bibnamefont{Sidery},
\bibinfo{author}{\bibnamefont{R.}} \bibnamefont{Sturani},
\bibinfo{author}{\bibnamefont{A.}} \bibnamefont{Vecchio},
\bibinfo{journal}{{ Phys.\,Rev.\,D}}
\textbf{85},~\bibinfo{journal}{082003}
(\bibinfo{year}{2012}).

\bibitem[{\citenamefont{Pozzo-et-al}(2012)}]{Poz_et_al_2012_2}
\bibinfo{author}{\bibnamefont{T.G.F.}} \bibnamefont{Li},
\bibinfo{author}{\bibnamefont{W.D.}} \bibnamefont{Pozzo},
\bibinfo{author}{\bibnamefont{S.}} \bibnamefont{Vitale},
\bibinfo{author}{\bibnamefont{C.V.D.}} \bibnamefont{Broeck},
\bibinfo{author}{\bibnamefont{M.}} \bibnamefont{Agathos},
\bibinfo{author}{\bibnamefont{J.}} \bibnamefont{Veitch},
\bibinfo{author}{\bibnamefont{K.}} \bibnamefont{Grover},
\bibinfo{author}{\bibnamefont{T.}} \bibnamefont{Sidery},
\bibinfo{author}{\bibnamefont{R.}} \bibnamefont{Sturani},
\bibinfo{author}{\bibnamefont{A.}} \bibnamefont{Vecchio},
\bibinfo{journal}{{ J.\,Phys.: Conf.\,Ser.}}
\textbf{363},~\bibinfo{journal}{012028}
(\bibinfo{year}{2012}).

\bibitem[{\citenamefont{Cornish-Sampson-Yunes-Pretorius}(2011)}]{Corn_Samp_Yun_Pret2011_GWtests}
\bibinfo{author}{\bibnamefont{N.}} \bibnamefont{Cornish},
\bibinfo{author}{\bibnamefont{L.}} \bibnamefont{Sampson},
\bibinfo{author}{\bibnamefont{N.}} \bibnamefont{Yunes},
\bibinfo{author}{\bibnamefont{F.}} \bibnamefont{Pretorius},
\bibinfo{journal}{{ Phys.\,Rev.\,D}}
\textbf{84},~\bibinfo{journal}{062003}
(\bibinfo{year}{2011}).

\bibitem[{\citenamefont{Pozzo-Veitch-Vecchio}(2011)}]{Pozzo_Veitch_Vecchio2011}
\bibinfo{author}{\bibnamefont{W.}} \bibnamefont{Del Pozzo},
\bibinfo{author}{\bibnamefont{J.}} \bibnamefont{Veitch},
\bibinfo{author}{\bibnamefont{A.}} \bibnamefont{Vecchio},
\bibinfo{journal}{{ Phys.\,Rev.\,D}}
\textbf{83},~\bibinfo{journal}{082002}
(\bibinfo{year}{2011}).

\bibitem[{\citenamefont{Sivia-Skilling}(2006)}]{Sivia}
\bibinfo{author}{\bibnamefont{D.S.}} \bibnamefont{Sivia},
\bibinfo{author}{\bibnamefont{J.}} \bibnamefont{Skilling},
\bibinfo{title}{{in {\it  Data Analysis: A Bayesian Tutorial, 2nd ed.} }}
(\bibinfo{year}{Oxford University Press, New York, 2006}).

\bibitem[{\citenamefont{Will}(1994)}]{Will_1994_BD}
\bibinfo{author}{\bibnamefont{C.M.}} \bibnamefont{Will},
\bibinfo{journal}{{ Phys.\,Rev.\,D}}
\textbf{50},~\bibinfo{journal}{6058-6067}
(\bibinfo{year}{1994}).

\bibitem[{\citenamefont{Will}(1998)}]{Will_1998_MG}
\bibinfo{author}{\bibnamefont{C.M.}} \bibnamefont{Will},
\bibinfo{journal}{{ Phys.\,Rev.\,D}}
\textbf{57},~\bibinfo{journal}{2061}
(\bibinfo{year}{1998}).

\bibitem[{\citenamefont{Arun-Will}(2009)}]{Arun_Will_2009}
\bibinfo{author}{\bibnamefont{A.}} \bibnamefont{Pai},
\bibinfo{author}{\bibnamefont{K.G.}} \bibnamefont{Arun},
\bibinfo{journal}{{ Classical Quantum Gravity}}
\textbf{26},~\bibinfo{journal}{155002}
(\bibinfo{year}{2009})

\bibitem[{\citenamefont{Keppel-Ajith}(2010)}]{Kep_Ajith2010_MG_Bound}
\bibinfo{author}{\bibnamefont{D.}} \bibnamefont{Keppel},
\bibinfo{author}{\bibnamefont{P.}} \bibnamefont{Ajith},
\bibinfo{journal}{{ Phys.\,Rev.\,D}}
\textbf{82},~\bibinfo{journal}{122001}
(\bibinfo{year}{2010}). 

\bibitem[{\citenamefont{Mirshekari-Yunes-Will}(2012)}]{Mirs-Yunes-Will_2012_LV}
\bibinfo{author}{\bibnamefont{S.}} \bibnamefont{Mirshekari},
\bibinfo{author}{\bibnamefont{N.}} \bibnamefont{Yunes},
\bibinfo{author}{\bibnamefont{C.M.}} \bibnamefont{Will},
\bibinfo{journal}{{ Phys.\,Rev.\,D}}
\textbf{85},~\bibinfo{journal}{024041}
(\bibinfo{year}{2012}).

\bibitem[{\citenamefont{Buonanno-et-al}(2009)}]{taylorf2}
\bibinfo{author}{\bibnamefont{A.}} \bibnamefont{Buonanno},
\bibinfo{author}{\bibnamefont{B.R.}} \bibnamefont{Iyer},
\bibinfo{author}{\bibnamefont{E.}} \bibnamefont{Ochsner},
\bibinfo{author}{\bibnamefont{Y.}} \bibnamefont{Pan},
\bibinfo{author}{\bibnamefont{B.S.}} \bibnamefont{Sathyaprakash},
\bibinfo{journal}{{ Phys.\,Rev.\,D}}
\textbf{80},~\bibinfo{journal}{084043}
(\bibinfo{year}{2009}).

\bibitem[{\citenamefont{Arun-et-al}(2005)}]{Arun_et_al_2005}
\bibinfo{author}{\bibnamefont{K.G.}} \bibnamefont{Arun},
\bibinfo{author}{\bibnamefont{B.R.}} \bibnamefont{Iyer},
\bibinfo{author}{\bibnamefont{B.S.}} \bibnamefont{Sathyaprakash},
\bibinfo{author}{\bibnamefont{Pranesh A.}} \bibnamefont{Sundararajan},
\bibinfo{journal}{{ Phys.\,Rev.\,D}}
\textbf{71},~\bibinfo{journal}{084008};
\bibinfo{journal}{{ Phys.\,Rev.\,D}}
\textbf{72},~\bibinfo{journal}{069903(E)}
(\bibinfo{year}{2005}).

\bibitem[{\citenamefont{Apostolatos}(1995)}]{Apostolatos_FF}
\bibinfo{author}{\bibnamefont{T.A.}} \bibnamefont{Apostolatos},
\bibinfo{journal}{{ Phys.\,Rev.\,D}}
\textbf{52},~\bibinfo{journal}{605}
(\bibinfo{year}{1995}).

\bibitem[{\citenamefont{Sathyaprakash-Dhurandhar}(1991)}]{filters_gw_binaries}
\bibinfo{author}{\bibnamefont{B.S.}} \bibnamefont{Sathyaprakash},
\bibinfo{author}{\bibnamefont{S.V.}} \bibnamefont{Dhurandhar},
\bibinfo{journal}{{ Phys.\,Rev.\,D}}
\textbf{44},~\bibinfo{journal}{3819}
(\bibinfo{year}{1991}).

\bibitem[{\citenamefont{Bender-Orszag}(1999)}]{bender_orszag}
\bibinfo{author}{\bibnamefont{C.M.}} \bibnamefont{Bender},
\bibinfo{author}{\bibnamefont{S.A.}} \bibnamefont{Orszag},
\bibinfo{title}{{in {\it Advanced Mathematical Methods for Scientists and Engineers I, Asymptotic Methods and Perturbation Theory}}}
(\bibinfo{year}{Springer, New York, 1999}).

\bibitem[{\citenamefont{Chatziioannou-Yunes-Pretorius}(2011)}]{Chat_Yun_Corn2012_ExtendedppE}
\bibinfo{author}{\bibnamefont{K.}} \bibnamefont{Chatziioannou},
\bibinfo{author}{\bibnamefont{N.}} \bibnamefont{Yunes},
\bibinfo{author}{\bibnamefont{N.}} \bibnamefont{Cornish},
\bibinfo{journal}{{ Phys.\,Rev.\,D}}
\textbf{86},~\bibinfo{journal}{022004}
(\bibinfo{year}{2012}).

\bibitem[{\citenamefont{Vallisneri-Yunes}(2013)}]{Vallisneri_Yunes_stealth_bias}
\bibinfo{author}{\bibnamefont{M.}} \bibnamefont{Vallisneri},
\bibinfo{author}{\bibnamefont{N.}} \bibnamefont{Yunes},
\bibinfo{journal}{{ Phys.\,Rev.\,D}}
\textbf{87},~\bibinfo{journal}{102002}
(\bibinfo{year}{2013}).

\bibitem[{\citenamefont{LIGO}(2015)}]{LIGO_noise}
\bibinfo{title}{{  \it Advanced  LIGO  anticipated  sensitivity  curves, }}
\bibinfo{journal}{{ LIGO Document}}
\bibinfo{journal}{T0900288-v3}

\bibitem[{\citenamefont{VIRGO-noise}(2015)}]{VIRGO_noise}
\bibinfo{author}{\bibnamefont{F.}} \bibnamefont{Acernese, et al.}
\bibinfo{author} \bibnamefont{(VIRGO)},
\bibinfo{journal}{{ Classical Quantum Gravity}}
\textbf{32},~\bibinfo{journal}{024001}
(\bibinfo{year}{2015}).

\bibitem[{\citenamefont{Vallisneri}(2008)}]{use_abuse_fisher}
\bibinfo{author}{\bibnamefont{M.}} \bibnamefont{Vallisneri},
\bibinfo{journal}{{ Phys.\,Rev.\,D}}
\textbf{77},~\bibinfo{journal}{042001}
(\bibinfo{year}{2008}).

\bibitem[{\citenamefont{Maggiore}(2007)}]{maggiore}
\bibinfo{author}{\bibnamefont{M.}} \bibnamefont{Maggiore},
\bibinfo{title}{{in {\it Gravitational Waves} }}
(\bibinfo{year}{Oxford University Press, New York, 2007, Vol.~1}).

\bibitem[{\citenamefont{Yunes-Hughes}(2010)}]{binary_pulsar_ppE}
\bibinfo{author}{\bibnamefont{N.}} \bibnamefont{Yunes},
\bibinfo{author}{\bibnamefont{S.A.}} \bibnamefont{Hughes},
\bibinfo{journal}{{ Phys.\,Rev.\,D}}
\textbf{82},~\bibinfo{journal}{082002}
(\bibinfo{year}{2010}).

\bibitem[{\citenamefont{Lyne-et-al}(2004)}]{binary_pulsar}
\bibinfo{author}{\bibnamefont{A.G}} \bibnamefont{Lyne, et al},
\bibinfo{journal}{{Science}}
\textbf{303},~\bibinfo{journal}{1153}
(\bibinfo{year}{2004}).

\bibitem[{\citenamefont{Yagi-Stein-Yunes-Tanaka}(2012)}]{QMG}
\bibinfo{author}{\bibnamefont{K.}} \bibnamefont{Yagi},
\bibinfo{author}{\bibnamefont{L.C}} \bibnamefont{Stein},
\bibinfo{author}{\bibnamefont{N.}} \bibnamefont{Yunes},
\bibinfo{author}{\bibnamefont{T.}} \bibnamefont{Tanaka},
\bibinfo{journal}{{ Phys.\,Rev.\,D}}
\textbf{85},~\bibinfo{journal}{064022}
(\bibinfo{year}{2012}). 


\bibitem[{\citenamefont{Yagi}(2012)}]{xray_binary}
\bibinfo{author}{\bibnamefont{K.}} \bibnamefont{Yagi},
\bibinfo{journal}{{ Phys.\,Rev.\,D}}
\textbf{86},~\bibinfo{journal}{081504}
(\bibinfo{year}{2012}). 

\bibitem[{\citenamefont{Healy-et-al}(2012)}]{Healy_2012}
\bibinfo{author}{\bibnamefont{J.}} \bibnamefont{Healy},
\bibinfo{author}{\bibnamefont{T.}} \bibnamefont{Bode},
\bibinfo{author}{\bibnamefont{R.}} \bibnamefont{Haas},
\bibinfo{author}{\bibnamefont{E.}} \bibnamefont{Pazos},
\bibinfo{author}{\bibnamefont{P.}} \bibnamefont{Laguna},
\bibinfo{author}{\bibnamefont{D.M.}} \bibnamefont{Shoemaker},
\bibinfo{author}{\bibnamefont{N.}} \bibnamefont{Yunes},
\bibinfo{journal}{{ Classical Quantum Gravity}}
\textbf{29},~\bibinfo{journal}{232002}
(\bibinfo{year}{2012}).

\bibitem[{\citenamefont{Mirshekari-Will}(2013)}]{Mirshekari_Will_2013}
\bibinfo{author}{\bibnamefont{S.}} \bibnamefont{Mirshekari},
\bibinfo{author}{\bibnamefont{C.M.}} \bibnamefont{Will},
\bibinfo{journal}{{ Phys.\,Rev.\,D}}
\textbf{87},~\bibinfo{journal}{084070}
(\bibinfo{year}{2013})

\bibitem[{\citenamefont{Yunes-Pani-Cardoso}(2012)}]{Yunes_Pani_Cardoso_2012}
\bibinfo{author}{\bibnamefont{N.}} \bibnamefont{Yunes},
\bibinfo{author}{\bibnamefont{P.}} \bibnamefont{Pani},
\bibinfo{author}{\bibnamefont{V.}} \bibnamefont{Cardoso},
\bibinfo{journal}{{ Phys.\,Rev.\,D}}
\textbf{85},~\bibinfo{journal}{102003}
(\bibinfo{year}{2012})

\bibitem[{\citenamefont{Zaglauer}(2012)}]{Zaglauer_1992}
\bibinfo{author}{\bibnamefont{H.W.}} \bibnamefont{Zaglauer},
\bibinfo{journal}{{ Astrophys.\,J.}}
\textbf{393},~\bibinfo{journal}{685-696}
(\bibinfo{year}{1992})

\bibitem[{\citenamefont{Yagi-Yunes-Tanaka}(2012)}]{Dyn_CS}
\bibinfo{author}{\bibnamefont{K.}} \bibnamefont{Yagi},
\bibinfo{author}{\bibnamefont{N.}} \bibnamefont{Yunes},
\bibinfo{author}{\bibnamefont{T.}} \bibnamefont{Tanaka},
\bibinfo{journal}{{Phys.~Rev.~Lett.}}
\textbf{109},~\bibinfo{journal}{251105}
(\bibinfo{year}{2012}). 


\bibitem[{\citenamefont{Beringer_et_al}(2012)}]{PDG_2012}
\bibinfo{author}{\bibnamefont{J.}} \bibnamefont{Beringer, et al. \,(Particle Data Group)},
\bibinfo{journal}{{ Phys.\,Rev.\,D}}
\textbf{86},~\bibinfo{journal}{010001}
(\bibinfo{year}{2012}).

\bibitem[{\citenamefont{Kosteleck\'{y}}(1998)}]{kostelec}
\bibinfo{author}{\bibnamefont{D.}} \bibnamefont{Colladay},
\bibinfo{author}{\bibnamefont{V.A.}} \bibnamefont{Kosteleck\'{y}},
\bibinfo{journal}{{ Phys.\,Rev.\,D}}
\textbf{58},~\bibinfo{journal}{116002}
(\bibinfo{year}{1998})

\bibitem[{\citenamefont{Kosteleck\'{y}}(2004)}]{kostelec_2004}
\bibinfo{author}{\bibnamefont{V.A.}} \bibnamefont{Kosteleck\'{y}},
\bibinfo{journal}{{ Phys.\,Rev.\,D}}
\textbf{69},~\bibinfo{journal}{105009}
(\bibinfo{year}{2004}).

\bibitem[{\citenamefont{Kosteleck\'{y}}(2015)}]{kostelec_2015}
\bibinfo{author}{\bibnamefont{V.A.}} \bibnamefont{Kosteleck\'{y}},
\bibinfo{author}{\bibnamefont{J.D.}} \bibnamefont{Tasson},
\bibinfo{journal}{{Phys.\,Lett.\,B}}
\textbf{749},~\bibinfo{journal}{551}
(\bibinfo{year}{2015}).

\bibitem[{\citenamefont{Berti-Gair-Sasana}(2011)}]{Bert-Gair-Ses_2011_MG}
\bibinfo{author}{\bibnamefont{E.}} \bibnamefont{Berti},
\bibinfo{author}{\bibnamefont{J.}} \bibnamefont{Gair},
\bibinfo{author}{\bibnamefont{A.}} \bibnamefont{Sesana},
\bibinfo{journal}{{ Phys.\,Rev.\,D}}
\textbf{84},~\bibinfo{journal}{101501}
(\bibinfo{year}{2011}).

\bibitem[{\citenamefont{Amendola-Charmousis-Davis}(2007)}]{cassini}
\bibinfo{author}{\bibnamefont{L.}} \bibnamefont{Amendola},
\bibinfo{author}{\bibnamefont{C.}} \bibnamefont{Charmousis},
\bibinfo{author}{\bibnamefont{S.C.}} \bibnamefont{Davis},
\bibinfo{journal}{{ J.\,Cosmol.\,Astropart.\,Phys.}}
\textbf{10},~\bibinfo{journal}{004}
(\bibinfo{year}{2007}).

\bibitem[{\citenamefont{Bertotti-Iess-Tortora}(2003)}]{cassini2}
\bibinfo{author}{\bibnamefont{B.}} \bibnamefont{Bertotti},
\bibinfo{author}{\bibnamefont{L.}} \bibnamefont{Iess},
\bibinfo{author}{\bibnamefont{P.}} \bibnamefont{Tortora},
\bibinfo{journal}{{ Nature}}
\textbf{425},~\bibinfo{journal}{374-376}
(\bibinfo{year}{2003}).

\bibitem[{\citenamefont{Pani-Berti-Cardoso-Read}(2011)}]{EDGB_bound1}
\bibinfo{author}{\bibnamefont{P.}} \bibnamefont{Pani},
\bibinfo{author}{\bibnamefont{E.}} \bibnamefont{Berti},
\bibinfo{author}{\bibnamefont{V.}} \bibnamefont{Cardoso},
\bibinfo{author}{\bibnamefont{J.}} \bibnamefont{Read},
\bibinfo{journal}{{ Phys.\,Rev.\,D}}
\textbf{84},~\bibinfo{journal}{104035}
(\bibinfo{year}{2011}).

\bibitem[{\citenamefont{Yunes_et_al}(2016)}]{test_gr_implications}
\bibinfo{author}{\bibnamefont{N.}} \bibnamefont{Yunes},
\bibinfo{author}{\bibnamefont{K.}} \bibnamefont{Yagi},
\bibinfo{author}{\bibnamefont{F.}} \bibnamefont{Pretorius},
\bibinfo{journal}{arXiv:1603.08955}
(\bibinfo{year}{2016}).

\bibitem[{\citenamefont{Kosteleck\'{y}}(2016)}]{test_gr_LV}
\bibinfo{author}{\bibnamefont{V.A.}} \bibnamefont{Kosteleck\'{y}},
\bibinfo{author}{\bibnamefont{M.}} \bibnamefont{Mewes},
\bibinfo{journal}{arXiv:1602.04782}
(\bibinfo{year}{2016}).


\bibitem[{\citenamefont{Allen}(1996)}]{Allen_1996}
\bibinfo{author}{\bibnamefont{B.}} \bibnamefont{Allen},
\bibinfo{journal}{arXiv:9607075}
(\bibinfo{year}{1996}).

\bibitem[{\citenamefont{Schutz}(2011)}]{Schutz_2011}
\bibinfo{author}{\bibnamefont{B.F.}} \bibnamefont{Schutz},
\bibinfo{journal}{{ Classical Quantum Gravity}}
\textbf{25},~\bibinfo{journal}{125023}
(\bibinfo{year}{2011}).

\bibitem[{\citenamefont{MTW}(1973)}]{MTW}
\bibinfo{author}{\bibnamefont{C.W.}} \bibnamefont{Misner},
\bibinfo{author}{\bibnamefont{K.S.}} \bibnamefont{Thorne},
\bibinfo{author}{\bibnamefont{J.A.}} \bibnamefont{Wheeler},
\bibinfo{title}{{in {\it Gravitation} }}
(\bibinfo{year}{W.H. Freeman, San Francisco, 1973}).

\bibitem[{\citenamefont{Goldstein}(1980)}]{goldstein_mech}
\bibinfo{author}{\bibnamefont{H.}} \bibnamefont{Goldstein},
\bibinfo{title}{{in {\it Classical Mechanics} }}
(\bibinfo{year}{Addison-Wesley, Reading, MA, 1980, 2nd Ed.}).

\bibitem[{\citenamefont{Isi-Weinstein-Mead-Pitkin}(2015)}]{CW_pol}
\bibinfo{author}{\bibnamefont{M.}} \bibnamefont{Isi},
\bibinfo{author}{\bibnamefont{A.J.}} \bibnamefont{Weinstein},
\bibinfo{author}{\bibnamefont{C.}} \bibnamefont{Mead},
\bibinfo{author}{\bibnamefont{M.}} \bibnamefont{Pitkin},
\bibinfo{journal}{{ Phys.\,Rev.\,D}}
\textbf{91},~\bibinfo{journal}{082002}
(\bibinfo{year}{2015})

\end{thebibliography}

\end{document}